\documentclass[12pt, aps, pre, preprint]{revtex4}
\usepackage{graphicx}
\usepackage{epsfig}
\usepackage{latexsym}
\usepackage{amssymb}
\usepackage{bm}
\usepackage{float}
\usepackage{subfigure}
\usepackage{amsmath}
\usepackage{amsfonts,color}
\usepackage{tabu}
\usepackage{array}
\usepackage{rotating}
\usepackage{multirow}
\newcolumntype{1}{>{\centering\arraybackslash}p{.75in}}
\newcolumntype{2}{>{\centering\arraybackslash}p{4.3cm}}
\newcolumntype{3}{>{\centering\arraybackslash}p{3cm}}
\newcolumntype{4}{>{\centering\arraybackslash}p{5.5cm}}
\newcolumntype{5}{>{\centering\arraybackslash}p{1.5in}}
\begin{document}

\title{Exploring extra dimensions with scalar fields}
\author{Katherine Brown$^{(1)}$, Harsh Mathur$^{(2)}$ and Mike Verostek$^{(1)}$}
\affiliation{$^{(1)}$Department of Physics, Hamilton College, Clinton, NY 13323}

\affiliation{$^{(2)}$Department of Physics, Case Western Reserve University, Cleveland, Ohio 44106-7079}
\begin{abstract}
This paper provides a pedagogical introduction to the physics of extra dimensions
by examining the behavior of scalar fields in three landmark models: the ADD,
Randall-Sundrum and DGP spacetimes. Results of this analysis provide
qualitative insights into the corresponding behavior of gravitational fields and elementary particles in each of these models.
In these ``brane world'' models the familiar four dimensional spacetime of everyday experience is called the
brane and is a slice through a higher dimensional spacetime called the bulk. The particles and fields of the standard
model are assumed to be confined to the brane while gravitational fields are assumed to propagate in the bulk.
For all three spacetimes we calculate the spectrum of propagating scalar wave modes and the scalar
field produced by a static point source located on the brane.
For the ADD and Randall-Sundrum models,
at large distances the field looks like that of a point source in four spacetime dimensions, but at short distances
it crosses over to a form appropriate to the higher dimensional spacetime. 
For the DGP model the field has the higher dimensional form at long distances rather than short. The behavior of these scalar fields, derived using only undergraduate level mathematics, closely mirror the results that one would obtain by performing the far more difficult task of analyzing the behavior of gravitational fields in these spacetimes.
\end{abstract}

\maketitle

\section{Introduction}

On the basis of everyday experience and on all scales that have been probed experimentally, it seems the space we inhabit is three dimensional. 
Nonetheless it remains a possibility that additional spatial dimensions exist.  A fourth spatial dimension
was invoked by Kaluza and Klein in the 1920s in order to develop a unified theory of electromagnetism
and gravity. Klein \cite{klein}  introduced the hypothesis that the extra dimension was rolled up, with an
estimated circumference of order $10^{-32}$ m, too small to be directly observed by experiments then and now. The idea of extra dimensions was revived in the 1980s 
by the discovery that superstring theory is only consistent in ten spacetime dimensions \cite{zwiebach}.
At first string theorists presumed that the extra dimensions were rolled up with a circumference on the Planck
scale ($\ell_P = 1.6 \times 10^{-35}$ m), and hence unobservable. However, several subsequent models challenged this notion.

In 1998 Arkani-Hamed, Dimopoulos, Dvali (ADD) and Antoniadis showed extra dimensions could be
much larger and yet remain hidden \cite{add, antoniadis}.  
Moreover ADD found that sufficiently large extra dimensions could explain the 
hierarchy problem, namely, the weakness of gravity compared to other forces of nature, or, equivalently,
the lightness of elementary particles compared to the Planck mass scale (which is inversely proportional
to the strength of gravity). ADD showed that large extra dimensions allow gravity to be
comparable in strength to other forces on short distance scales but to appear much weaker on 
scales that are large compared to the extra dimension.
Randall and Sundrum (RS) then showed that a warped extra dimension provided a different 
explanation 
of the hierarchy problem
\cite{rs1}; remarkably since their extra dimension was warped it could be
small and yet explain the hierarchy problem \cite{rs1}, or infinite and yet remain hidden \cite{rs2}. 
Somewhat later Dvali, Gabadadze and Porrati (DGP) invented another model with an infinite
extra dimension; however unlike ADD and RS, the DGP model sought to explain the observed accelerated expansion of the Universe 
 \cite{dgp}. 
In all of these models the familiar four dimensional
spacetime inhabited by particles and fields of the standard model is a brane: a slice through 
a higher dimensional universe called the bulk. 

Brane worlds with extra dimensions have engendered tremendous public
interest. Randall's {\em Warped Passages} \cite{randall} was a bestseller and a version of the Randall Sundrum
model was the scientific basis of the five dimensional world depicted in Christopher Nolan's
2014 movie {\em Interstellar} \cite{thorne}.
Yet there remains a gap between popular accounts like \cite{randall, thorne, krauss}, and review articles, such as \cite{hewett, particledata, claudia, sundrum}, that require advanced graduate
coursework in general relativity, field theory and standard model physics. The purpose of this article is to provide a pedagogical introduction to this subject at a level accessible to
undergraduate physics majors.  The target audience for our paper
includes physicists in other fields as well as undergraduate and graduate students. No specialized
background is required beyond one semester courses in electromagnetism and quantum mechanics. 
Indeed the material in this paper could be incorporated into such courses as well as introductory 
courses on particle physics or general relativity.

We show that a simple scalar field serves as a proxy for 
gravitational fields and elementary particles in the 
ADD, RS, and DGP models. For each of the three models
we calculate the solutions to the scalar wave equation
in the absence of sources, and the scalar 
field for a point source located on the brane. For the
ADD and RS models we find that on long distance scales
the field of a point source has the form appropriate to three spatial
dimensions; the form appropriate for higher dimensions 
becomes manifest only on short distance scales. For
the DGP model by contrast the field of a point source
has the form appropriate to three dimensions on short
distance scales and the higher dimensional form becomes
manifest on long distance scales. The behavior of the
field due to a point source reveals some of the most compelling
features of these models, in particular, their solutions to
the hierarchy and dark energy problems. We show that these features can be understood without invoking the machinery of tensor calculus or differential geometry required to fully analyze the gravitational field. It is not surprising that scalar fields serve as a useful caricature for more complicated fields, for example in electromagnetism courses it is common practice to introduce scalar waves as a prelude to studying electromagnetic waves \cite{feynman, griffiths}. It is also not surprising that the spatial dependence of the field of a point source changes when the number of spatial dimensions changes: Gauss' law for the gravitational or electrostatic field implies that in $d$ spatial dimensions the field must fall off like $1/r^{d-1}$, so in three spatial dimensions we get the famous $1/r$ behavior but in four spatial dimensions the fall off goes like $1/r^2$.

We begin in Section II with the familiar
four dimensional Minkowski spacetime; we solve the scalar wave equation and calculate the field 
of a static point source therein. These results lay the groundwork and provide insights into the physics
of gravitational fields and elementary particles. In Sections III, IV and V we then carry out a parallel analysis 
in the higher dimensional spacetimes of the ADD, RS and DGP models. Within each section, we identify the same constructs for each model in turn, labeled with italicized paragraph headings. The geometry of each spacetime is described; the necessary differential geometry background is presented in an appendix \cite{som}.  The hierarchy problem and its resolution within
the ADD and RS models are explained, as are the cosmological implications of the DGP model. In Section VI
we conclude by briefly discussing the experimental implications and tests of the three models. 
It is worth noting here that the simplest versions of the ADD and DGP models discussed in this paper are
ruled out by data; however descendants of these models as well as the RS model itself remain viable. Key results 
presented in the paper
are summarized in Table 1. 
In order to make our article more useful for students and instructors we provide a primer on differential geometry in Appendix A, on one dimensional quantum mechanics in Appendix B, and a number of problems in Appendix C; 
the appendices are available as supplementary online material \cite{som}. 

\section{Massless Scalar Field Theory in Minkowski Spacetime}
\label{minksection}
 A. {\em The spacetime interval.} Minkowski spacetime is the familiar spacetime from special relativity. 
The interval between two neighboring events, with coordinates $(t,x,y,z)$ and $(t+dt,x +dx, y+dy, z+dz, )$,
is given by 
\begin{equation}
ds^2 = dt^2 - dx^2 - dy^2 - dz^2
\label{eq:minkowskimetric}
\end{equation}
where $-\infty < t,x, y, z < \infty$ and we choose a system
of units wherein the speed of light $c = 1$.

B. {\em The wave equation.} A scalar wave propagating in 
Minkowski spacetime obeys the familiar wave equation
\begin{equation}
\Box^2 \phi = 
\left(
\frac{\partial^2}{\partial t^2} - 
\frac{\partial^2}{\partial x^2} -
\frac{\partial^2}{\partial y^2} -
\frac{\partial^2}{\partial z^2} \right) \phi = s ({\mathbf r}, t).
\label{eq:minkwave}
\end{equation}
where $\phi$ is the scalar field and $s$ is its source.\\

\indent C.{\em Source-free solution.} If $s=0$,
 the wave equation (\ref{eq:minkwave}) has plane wave solutions
$\phi = f({\mathbf r}; {\mathbf k}) \exp (-i \omega t )$ where 
\begin{equation}
f( {\mathbf r}; {\mathbf k}) = \frac{1}{(2 \pi)^{3/2}} \exp ( i {\mathbf k} \cdot {\mathbf r} )
\label{eq:planemode}
\end{equation}
and the frequency $\omega$ and wave-vector ${\mathbf k}$ obey the
dispersion relation $\omega = k$. We refer to the function $f({\mathbf r}; {\mathbf k})$ as the mode function. 

\indent D. {\em Static point source solution.}
Consider a static point  source $s({\mathbf r}) = \lambda \delta ({\mathbf r})$; more general static
sources can be analyzed by superposition. 
 For a static source
the field $\phi$ will also be independent of time. 
Thus
we are just solving Poisson's equation with a delta function source and so the answer can immediately
be written down from the known electrostatic potential of a point charge. 
However for later use it is helpful to describe another approach. 
Taking the Fourier transform 
of eq (\ref{eq:minkwave}) yields 
$k^2 \tilde{\phi}({\mathbf k}) = \lambda$ where $\tilde{\phi}({\mathbf k})$ is the spatial Fourier
transform of $\phi({\mathbf r})$. Inverting the Fourier transform yields \footnote{The Fourier integral is most easily evaluated by working in spherical coordinates and integrating over the angular variables first. The radial integral
can be performed making use of $\int_0^\infty dx \; (\sin x)/x = \pi/2$.} 
\begin{equation}
\phi ({\mathbf r}) = \lambda \int \frac{ d {\mathbf k} }{ (2 \pi)^3} \; 
\frac{1}{k^2} \exp ( i {\mathbf k} \cdot {\mathbf r} ) = \frac{\lambda}{4 \pi r}.
\label{eq:fouriercoulomb}
\end{equation}

Equations (\ref{eq:planemode}) and (\ref{eq:fouriercoulomb}) are analogous to familiar results from electrostatics and gravity:  in the absence of sources
Maxwell's equations and Einstein's equations also have plane wave solutions with the same dispersion
$\omega = k$, 
and a static point charge or point mass  produces a $1/r$ potential. Thus we see that solutions to the scalar field equations provides insights into the solutions
to Maxwell's equations and Einstein's equations as well.

It is useful here to briefly
discuss a variant on the scalar wave equation in  
Minkowski spacetime, called the Klein-Gordon model or massive scalar field theory.
In this model the scalar field obeys 
\begin{equation}
\Box^2 \phi + \mu^2 \phi = s.
\label{eq:kleingordon}
\end{equation}
where the mass parameter $\mu$ has units of inverse length. 
We may regard the conventional wave eq (\ref{eq:minkwave}) as the massless limit 
($\mu \rightarrow 0$) 
of the Klein-Gordon model. 
In the absence of sources, plane waves are still solutions,
with $f$ given by eq (\ref{eq:minkwave}), but the frequency and wave-vector are related by the dispersion relation
$\omega = \sqrt{k^2 + \mu^2}$.

A key insight is that in quantum field theory the
Klein-Gordon model describes 
massive particles. This can be motivated by multiplying both sides of the dispersion relation by
$\hbar$ and recalling that $\hbar \omega$ is the energy and $\hbar k$ the momentum by Planck's
formula and the de Broglie relation respectively. Then comparing the dispersion relation 
to the energy-momentum relation 
of a particle of mass $m$ in special relativity reveals that the Klein-Gordon model describes 
a particle of mass $m = \mu \hbar/c$, where we have momentarily restored the factors of $c$.
The wave equation corresponds to $\mu = 0$ and describes a massless particle in quantum
field theory, the scalar field 
analog of the photon and the graviton. 

In the Klein-Gordon model also $\phi$ will 
be time independent  for a static source.
And again it is sufficient to analyze a point
source $s = \lambda \delta ({\mathbf r})$ and obtain more general static solutions by
superposition. Taking the Fourier transform with respect to the spatial dependence in 
eq (\ref{eq:kleingordon}) yields 
$(k^2 + \mu^2) \tilde{\phi} = \lambda$. 
Inverting the Fourier transform yields the Yukawa potential
\begin{equation}
\phi ({\mathbf r}) = \lambda \int \frac{ d {\mathbf k} }{ (2 \pi)^3 } \frac{1}{ k^2 + \mu^2 } \exp ( i {\mathbf k} 
\cdot {\mathbf r} ) = \frac{ \lambda }{ 4 \pi r} \exp ( - \mu r ).
\label{eq:yukawa}
\end{equation}
At short distances, $r \ll \mu^{-1}$, the Yukawa potential reproduces the massless
$1/r$ behavior but at long distances, $r \gg \mu^{-1}$, the Yukawa potential vanishes exponentially. 
We will see later that the exponential disappearance of the field is the key to hiding the extra dimensions
in the ADD model and in one version of the RS model as well.

It is a good exercise to verify eq (\ref{eq:yukawa}). 
The Fourier integral is most easily evaluated by working in spherical coordinates and 
integrating over the angular variables first. The radial integral can be performed by making use of
the definite integral 
\begin{equation}
\int_0^\infty d k \; \frac{k \sin k r}{k^2 + \mu^2} = \frac{\pi}{2} \exp( - \mu r).
\label{eq:radialintegralthisone}
\end{equation}
Eq (\ref{eq:radialintegralthisone}) can be obtained by contour integration or by more elementary means
\footnote{Evaluate the Fourier transform
of $f(x)$ defined as $f(x) = \exp( - \mu x )$ for $x > 0$ and 
$f(x) = - \exp ( - \mu x )$ for $x<0$. This is readily found to be $-2ik/(k^2 + \mu^2)$. Now write $f$
in terms of its Fourier transform, take the real part of both sides, and note that the Fourier
integral over $k$ is even.}.

\section{The ADD Model}
A major motivation for considering models with extra dimensions is that they offer an elegant solution to
the `hierarchy problem'. The hierarchy problem can be simply understood in terms of dimensional analysis.
From the three fundamental constants $c, G$ and $\hbar$ one can form only one expression with units of 
energy, the Planck energy, given by $E_P = \sqrt{ \hbar c^5/ G} = 10^{28}$eV\footnote{This is a stupendous 
energy scale. For comparison the energy at which protons are collided by the Large Hadron collider 
is $\sim$10$^{13}$eV. The highest energy subatomic particles detected are ultra high energy 
cosmic rays which have an energy of order $10^{21}$eV.}. 
Assuming that these are the 
only parameters that should appear in a fundamental theory of nature it follows that
$E_P$ should be the characteristic energy scale of that fundamental theory. In the standard model
of particle physics an important energy scale is the electroweak scale $E_{{\rm EW}}$ 
at which the electromagnetic and weak
nuclear forces are unified. In a fundamental theory that
subsumes the standard model, the dimensional argument suggests that 
the electroweak scale $E_{ EW}$ should be of the
same order as the Planck scale $E_P$. But in fact the electroweak scale $E_{ EW} \approx 10^{13}$eV $\ll E_P$. The vast
discrepancy between these scales is the hierarchy problem. The problem grows more acute in context of quantum
field theory wherein such a large discrepancy in these scales is unstable. (For an authoritative non-technical discussion 
placing the hierarchy problem in context of quantum field theory see ref \cite{randall}.) Also note that 
as $G \rightarrow 0$, $E_P \rightarrow \infty.$ The weaker gravity is, the worse the hierarchy problem becomes.
Thus we may regard the hierarchy problem as a measure of the weakness of gravity compared to other
forces. Any theory that seeks to explain the hierarchy problem must explain the vast gulf in the Planck and 
electroweak scales; the ADD and RS models do this in different ways by
incorporating extra dimensions. 

 A. {\em The spacetime interval. } The ADD model can include any number of extra spatial dimensions; here we consider the simplest version in which  there are four spatial dimensions and one temporal dimension.
The four spatial dimensions are specified by Cartesian coordinates $(x, y, z, w)$ where the extra dimension $w$ is rolled up with a circumference $L$. More precisely, $(x, y, z, w)$ and $(x, y, z, w+L)$ correspond to the same spatial point. 
The spacetime interval between two nearby events located 
at $(t, x, y, z, w)$ and $(t + d t, x + dx, y + d y, z + d z, w + d w)$ is given by
\begin{equation}
ds^2 = d t^2 - dx^2 - dy^2 - dz^2 - dw^2.
\label{eq:addmetric}
\end{equation}
The four dimensional cross-section of ADD spacetime 
corresponding to $w = 0$ is called the brane. 
More generally branes are lower dimensional hyper-surfaces in a higher dimensional
spacetime.
Branes are assumed to trap standard model particles
and fields so that they effectively only propagate on the brane. (A useful analogy to a particle trapped on the brane is a water droplet on a shower curtain \cite{randall}.  Although the droplet lives in a three dimensional world, it is confined to travel in only two dimensions.) 
In contrast, the gravitational field propagates throughout the
five dimensional spacetime, called the bulk.

B. {\em The Wave Equation}. By analogy to eq (\ref{eq:minkwave}) 
the scalar wave equation in ADD spacetime has the form
\begin{equation}
\Box^2 \phi = \left(
\frac{\partial^2}{\partial t^2} -
\frac{ \partial^2 }{\partial x^2} -
\frac{ \partial^2 }{\partial y^2} -
\frac{ \partial^2 }{\partial z^2} -
\frac{ \partial^2 }{\partial w^2} \right) \phi = s ({\mathbf r}, t).
\label{eq:fivedlapacian}
\end{equation}
where $\phi$ is the scalar field and $s$ is its source. 

C. {\em Source-free solution: zero mode and Kaluza-Klein modes}. In the absence of sources the wave
equation has plane wave solutions $\phi = f({\mathbf r}, w; {\mathbf k}, \kappa_n) \exp (-i \omega t )$
where the mode function
\begin{equation}
f({\mathbf r}, w; {\mathbf k}, \kappa_n) = \frac{1}{(2 \pi)^{3/2}} \frac{1}{\sqrt{L}} \exp ( i {\mathbf k} \cdot {\mathbf r} )
\exp ( i \kappa_n w )
\label{eq:addwave}
\end{equation}
and the frequency $\omega$ obeys the dispersion relation 
\newline
$\omega = \sqrt{ k^2 + \kappa_n^2 }$. 
In order to ensure the periodicity $f( {\mathbf r}, w + L ; {\mathbf k}, \kappa_n )= f( {\mathbf r}, w; 
{\mathbf k}, \kappa_n )$ the extra dimensional component of the wave-vector $\kappa_n$ obeys
the quantization 
\begin{equation}
\kappa_n = \frac{2 \pi n}{L}
\label{eq:addquantization}
\end{equation}
where $n = \ldots, -1, 0, 1, 2, \ldots$ is an integer. 
Waves with $n=0$ are called zero modes. Zero modes behave essentially like waves in 
four dimensional spacetime. They are labelled by a wave vector ${\mathbf k}$ that only
has components along the brane and they have the massless dispersion $\omega = k$. Modes with non-zero $n$ are 
called Kaluza-Klein modes. Kaluza-Klein modes have 
an $\omega$-${\mathbf k}$ dispersion corresponding to the massive wave equation
with a mass parameter $\mu \rightarrow \kappa_n$. 

D. {\em Static point source solution.} 
Next let us examine the solution to the wave equation (\ref{eq:addwave})
for a static source that is confined to the brane. As in Section \ref{minksection} we consider a point source 
$s = \lambda \delta ({\mathbf r}) \delta( w)$;
more complicated distributions can be treated by superposition. 
The resulting field $\phi$ will be time independent and has the Fourier expansion
\begin{equation}
\phi( {\mathbf r}, w ) = \sum_{n=-\infty}^\infty \int d {\mathbf k} \; \tilde{\phi}( {\mathbf k}, \kappa_n) 
f ({\mathbf r}, w; {\mathbf k}, \kappa_n)
\label{eq:addpoint}
\end{equation}
where the Fourier amplitudes $\tilde{\phi}({\mathbf k}, \kappa_n)$ are given by
\begin{equation}
\tilde{\phi} ({\mathbf k}, \kappa_n) = \int d {\mathbf r} \; \int_0^L d w \; 
\phi( {\mathbf r}, w ) f^\ast ({\mathbf r}, w; {\mathbf k}, \kappa_n).
\label{eq:addfourieranalysis}
\end{equation}
In Fourier space the wave eq (\ref{eq:addwave}) takes the form
\newline
$(k^2 + \kappa_n^2) \tilde{\phi} = \lambda/\sqrt{L} (2 \pi)^{3/2}$ and hence
\begin{equation}
\phi ({\mathbf r}, w) = \frac{1}{L} \sum_{n=-\infty}^{\infty} \int \frac{d {\mathbf k}}{(2 \pi)^3} \;
\frac{\lambda}{k^2 + \kappa_n^2} \exp ( i {\mathbf k} \cdot {\mathbf r} ) \exp ( i \kappa_n w ).
\label{eq:intermediatepointadd}
\end{equation} 
The ${\bf k}$ integral can be performed by comparison to eq (\ref{eq:yukawa}). Hereafter
let us consider only the value of the field on the brane by setting $w = 0$. We obtain
\begin{equation}
\phi({\mathbf r}, 0) = \frac{1}{L} \sum_{n = - \infty}^{\infty} 
\frac{\lambda}{4 \pi r}
\exp \left( - \frac{2 \pi | n  | r}{L}  \right)
\label{eq:addpointsolution}
\end{equation}
Eq (\ref{eq:addpointsolution}) is the field of a point source that is located on the brane at the
origin at a distance $r$ from the source. The sum over modes, $n$, can be performed 
exactly but it is more illuminating to consider limiting cases. For $r \gg L$ only the zero mode
($n = 0$) contributes to the field---the contribution of the Kaluza-Klein modes is 
exponentially suppressed. We obtain
\begin{equation}
\phi ({\mathbf r}, 0) \approx \frac{\lambda}{L} \frac{1}{4 \pi r}
\label{eq:oneoverr}
\end{equation}
At distances that are long compared to the size of the extra dimension we obtain
the $1/r$ behavior characteristic of four spacetime dimensions; the extra dimension is
essentially invisible. On the other hand for $r \ll L$ all modes contribute and we obtain
\begin{equation}
\phi( {\mathbf r}, 0 ) \approx \frac{ \lambda}{4 \pi^2} \frac{1}{r^2}.
\label{eq:oneoverrsquare}
\end{equation}
At distances that are short compared to the extra dimension we obtain the $1/r^2$ behavior
characteristic of five spacetime dimensions. 
An intuitive explanation of these limiting cases is provided by examining the field lines
of a point source in ADD spacetime (see chapter 4 of ref \cite{thorne}).

{\em Hierarchy revisited.} Standard model fields are confined to the brane in the ADD model.
Only gravity propagates through the bulk.
Above we have analyzed a scalar field that propagates
in the bulk as a proxy for the gravitational field. 
Now we demonstrate the essence of the solution to the hierarchy problem within the ADD model.

Suppose we were interested in the gravitational
field of a point mass $m$ that lives on the brane. Based on the above analysis, at short distances we expect that the Newtonian
potential will have the five dimensional form $G_5 m/ r^2$ where $G_5$ is characterizes the gravitational interaction, the five dimensional version of the Newtonian constant $G$.
On the other hand at long distances we expect the Newtonian potential would have the familiar four dimensional
form $G m/r$.
Comparing these forms to eqs (\ref{eq:oneoverr})
and (\ref{eq:oneoverrsquare}) we infer that $\lambda = 4 \pi^2 G_5 m = 4 \pi G L m$. Hence we arrive at the important
result
\begin{equation}
G_5 =  \frac{GL}{\pi}.
\label{eq:hierarchysolved}
\end{equation}
Note that except for the factor of $\pi$ the form of eq (18) 
follows simply because the dimensions of $G_5$ have an extra
factor of length compared to the dimensions of $G$.
Eq (\ref{eq:hierarchysolved}) lies at the heart of the ADD solution to the hierarchy problem. 
The idea is that the fundamental Planck scale should be determined by $c, \hbar$ and $G_5$.  Simple dimensional analysis shows that the
Planck scale should be given by $E_{P5} = ( \hbar^2 c^6/G_5)^{1/3}$. 
In this expression we replace $G_5 \rightarrow GL$, in accordance with eq (\ref{eq:hierarchysolved}), 
and impose the requirement that $E_{P5} \sim 1$TeV (the electroweak scale) to obtain $L \sim 10^{13}$m.
Thus the hierarchy problem
is solved if we posit the existence of an extra dimension of this enormous size. 
Physically the idea is that gravity is not weaker than other forces; it merely appears weaker
on long length scales because gravitational fields spread out into the higher dimensional bulk
while other force fields remain concentrated on the brane. 

Similarly if there were two large extra dimensions we would have $G_6 \sim G L^2$. 
On dimensional grounds, the Planck scale would be given by $(\hbar^3 c^7/G_6)^{1/4}$. Replacing
$G_6 \rightarrow G L^2$ and imposing the requirement that $E_P \sim 1$TeV yields $L \sim 1$ mm.
Thus the six dimensional ADD model solves the hierarchy problem if we posit the existence of two 
extra dimensions on the millimeter scale (which is still enormous compared to the Planck length scale $\ell_P$). 
See section \ref{sec:discussion} for further discussion of the experimental implications of the ADD model
and the current experimental constraints on it. 

\section{The RS model}

There are two versions of the RS model known as RS1 \cite{rs1} and RS2 \cite{rs2}.
Here for simplicity we will focus primarily on the RS2 model and relegate the analysis of RS1
to a problem (see Appendix C \cite{som}). It is worth noting at the outset that in both
RS models there is exactly one extra spatial dimension whereas in the ADD model the number of
extra dimensions is arbitrary.
Another key difference between the two models is that 
the RS1 and RS2 spacetimes are curved whereas the ADD spacetime is flat.
Readers who do not have a background
in general relativity 
and curved spacetime 
may find it useful to read Appendix A \cite{som} before proceeding with the remainder of this section.

A. {\em The spacetime interval.}  An event in the RS2 
spacetime is specified by  $(t, x, y, z, w)$. where $0 \leq w < \infty$ and $- \infty < t, x, y, z < \infty$. 
The brane is the four dimensional cross section of this spacetime corresponding to $w = 0$.
The spacetime interval between neighboring
events located at $(t, x, y, z, w)$ and $(t + d t, x + d x, y + d y, z + d z, w + d w)$ is given by
\footnote{Randall and Sundrum did not merely assume this form but rather showed that it was 
compatible with general relativity if they made particular assumptions about the brane. 
However that analysis is beyond the scope of this article.} 
\begin{equation}
ds^2 = e^{-2 \gamma w} (dt^2 - dx^2 - dy^2 - dz^2) - dw^2.
\label{eq:rsmetric}
\end{equation}

$\gamma$ is a parameter with units of inverse length. 
For fixed $w$ the spacetime is essentially flat 
Minkowski spacetime but with an overall $w$ dependent ``warp factor'' $e^{-2 \gamma w}$. 
There is no system of coordinates in which the
spacetime interval has the flat form eq (\ref{eq:addmetric}) globally.

Although in principle we could continue to 
work with the coordinates $(t, x, y, z, w)$, it is more convenient to exchange the coordinate $w$ with  $\zeta$, where
$\zeta = e^{\gamma w}/\gamma$. This gives $\gamma^{-1} \leq \zeta < \infty$ and now the brane is  
the four dimensional cross section with 
$\zeta = \gamma^{-1}$.  In terms of these coordinates the spacetime interval in (\ref{eq:rsmetric}) is given by
\begin{equation}
ds^2 = \frac{1}{\gamma^2 \zeta^2} \left( dt^2 - dx^2 - dy^2 - dz^2 - d\zeta^2 \right)
\label{eq:conformal}
\end{equation}
as the diligent reader should verify. Eq (\ref{eq:conformal}) reveals that the RS2 spacetime
has the same interval as the ADD spacetime eq (\ref{eq:addmetric}) up to an overall factor. Since the two
spacetime intervals are related by an overall multiplicative factor, they are said to be conformally equivalent.
The coordinates $(t, x, y, z, \zeta)$ are called conformal coordinates. 

B. {\em The wave equation}. In conformal coordinates the scalar wave equation has the form
\begin{equation}
- \frac{\partial^2}{\partial \zeta^2} \phi + \frac{3}{\zeta} \frac{\partial}{\partial \zeta} \phi -  
\left( \frac{\partial^2}{\partial x^2} + \frac{\partial^2}{\partial y^2} + \frac{\partial^2}{\partial z^2} \right) \phi
+ \frac{\partial^2}{\partial t^2} \phi = \gamma^2 \zeta^2 s.
\label{eq:rswaveequation}
\end{equation}

Here $\phi$ is the scalar field and $s$ is the source. 
The reader should verify that eq (\ref{eq:rswaveequation}) follows from the spacetime interval 
eq (\ref{eq:conformal}) following the methods of appendix A \cite{som}. 

\indent C. {\em Source-free solution: zero mode and Kaluza-Klein modes}. The form of eq (\ref{eq:rswaveequation})
suggest that the solutions in the absence of sources will have the form of plane waves modulated
by a $\zeta$ dependent factor $\upsilon(\zeta)$, namely, 
$\phi = \upsilon(\zeta) \exp( i {\mathbf k} \cdot {\mathbf r} ) 
\exp (-i \omega t)$. Substitution of this ansatz in eq (\ref{eq:rswaveequation}) reveals that the modulation
factor obeys
\begin{equation}
\frac{d^2 \upsilon}{d \zeta^2} - \frac{3}{\zeta} \frac{d \upsilon}{d \zeta} + (\omega^2 - k^2) \upsilon = 0.
\label{eq:separationanxiety}
\end{equation}
We will see below that $\omega^2 \geq k^2$; hence it is convenient to write $\omega^2 - k^2$ as $ \mu^2$
in eq (\ref{eq:separationanxiety}). One solution to eq (\ref{eq:separationanxiety}) that is immediately
apparent is to take $v$ to be a constant independent of $\zeta$ if we also choose the massless dispersion
$\omega = k$. This solution is the zero-mode. It is an acceptable solution provided we impose 
Neumann boundary conditions $d v / d \zeta = 0$ on the brane.

{\em Schr\"{o}dinger analogy.} Now we systematically construct all solutions 
to eq (\ref{eq:separationanxiety}). To this end, following \cite{rs2} we first transform to the independent
variable 
$\psi$ defined by $\upsilon = \zeta^{3/2} \psi$. 
Eq (\ref{eq:separationanxiety}) then has the form
\begin{equation}
- \frac{1}{2} \frac{d^2}{d \zeta^2} \psi + \frac{15}{8} \frac{1}{\zeta^2} \psi = \frac{\mu^2}{2} \psi.
\label{eq:schrodingeranalogy}
\end{equation}
Eq (\ref{eq:schrodingeranalogy}) coincides with the time-independent
Schr\"{o}dinger equation for a particle that is confined to the
half line $\zeta> \gamma^{-1}$ and that experiences a potential barrier $15/8\zeta^2$. In this analogy
the ``energy'' of the particle is $\mu^2/2$. Thus we can draw upon our intuition from
non-relativistic quantum mechanics in order to deduce the solutions to eq (\ref{eq:schrodingeranalogy}). 
The form of the potential suggests that there should be a continuum of states with $\mu^2 > 0$. These 
states constitute the Kaluza-Klein modes of the RS2 model. In addition there is a single
bound state which corresponds to the zero-mode of the RS2 model.

{\em Zero mode.} 
It may seem surprising that eq (\ref{eq:schrodingeranalogy}) has a bound state
solution since the potential is purely repulsive. However we still have the freedom to choose the boundary
condition we apply to the ``wave-function'' $\psi$ at the location of the brane $\zeta= \gamma^{-1}$. The boundary 
condition will have the form $\psi (\gamma^{-1}) = \lambda \psi'(\gamma^{-1})$ where $\lambda$ is a parameter that we 
have freedom to specify. The boundary
condition effectively functions like a contact interaction which can be repulsive or attractive \footnote{To understand
how a boundary condition is essentially a contact interaction consider a particle moving in one dimension
free except for a delta function potential at the origin $\lambda^{-1} \delta (x)$. If we look for solutions that are
symmetric about the origin we need to impose the boundary condition $\psi'(0) = \lambda^{-1} \psi(0)$. (This
can be derived by integrating the Schr\"{o}dinger equation across the origin as discussed in textbook treatments of the
delta function model).}. By choosing
the right boundary condition we can ensure that there will be a bound state with $\mu = 0$. Indeed if we 
set $\mu =0$ it is easy to see that eq (\ref{eq:schrodingeranalogy}) has power law solutions of the form
$\psi = \zeta^\beta$ with $\beta = -\frac{3}{2}$ or $\beta = \frac{5}{2}$. We discard the latter on the grounds that it is not
normalizable. Normalizing the former we conclude that eq (\ref{eq:schrodingeranalogy}) has the 
bound state solution with $\mu = 0$ given by 
\begin{equation}
\psi_0 = \frac{\sqrt{2}}{\gamma} \zeta^{-3/2}
\label{eq:rszeromode}
\end{equation} 
that satisfies the normalization condition $\int_{\gamma^{-1}}^\infty d \zeta \; | \psi_0 |^2 = 1$. 
The solution (\ref{eq:rszeromode}) satisfies
the boundary condition $\psi(\gamma^{-1}) = \lambda \psi'(\gamma^{-1})$ with $\lambda = - \frac{2}{3} \gamma^{-1}$ 
and hence this is the boundary condition we will impose on the scalar field at the brane
hereafter \footnote{Recall that $\upsilon = \zeta^{3/2} \psi$ and $\phi = \upsilon(\zeta) \exp(i {\mathbf k} \cdot {\mathbf r} )
\exp(-i \omega t)$. Transforming the boundary condition on $\psi$ into a boundary condition on $\upsilon$ we see
that $\phi$ simply satisfies Neumann boundary conditions on the brane, $\partial \phi/\partial \zeta |_{\zeta \rightarrow
\gamma^{-1}} = 0$.}. Undoing the transformation $v = \zeta^{3/2} \psi$ we see that $v_0 = \sqrt{2}/\gamma$,
which matches the form for the zero mode we had already deduced following eq (\ref{eq:separationanxiety}).

{\em Kaluza-Klein modes.} Determination of the positive energy continuum of solutions 
to eq (\ref{eq:schrodingeranalogy}) is a straightforward exercise in one dimensional
quantum mechanics. We relegate the details to appendix B \cite{som} and only quote the results here.
The solutions have the form
\begin{equation}
\psi (\zeta ; \mu) = \sqrt{\zeta} \left[
\alpha ( \mu ) J_2 ( \mu \zeta ) + \beta (\mu) Y_2 ( \mu \zeta ) \right].
\label{eq:rscontinuum}
\end{equation}
Here $J_2$ and $Y_2$ denote the Bessel and Neumann functions of second order. 
Eq (\ref{eq:rscontinuum}) is a solution to eq (\ref{eq:schrodingeranalogy}) for arbitrary 
coefficients $\alpha$ and $\beta$. The boundary condition on the brane at $\zeta = \gamma^{-1}$
fixes the ratio of $\alpha$ to $\beta$. We find
\begin{equation}
\alpha (\mu) = A Y_1 \left( \frac{ \mu }{\gamma} \right) \hspace{3mm}
{\rm and} \hspace{3mm} 
\beta (\mu) = - A J_1 \left( \frac{ \mu }{\gamma} \right).
\label{eq:alphabeta}
\end{equation}

We choose the overall constant $A$ by imposing the condition that $\alpha^2 + \beta^2 = \mu$. As shown
in appendix B \cite{som} this choice ensures that the continuum solutions satisfy the normalization condition
\begin{equation}
\int_{\gamma^{-1}}^\infty d \zeta \; \psi^\ast ( \zeta; \mu ) \psi (\zeta; \mu') = \delta (\mu - \mu').
\label{eq:continuumnormrs}
\end{equation}
Recall that solutions to the Schr\"{o}dinger equation constitute a complete set. For later use it is worth noting that 
the continuum solutions 
derived here together with the zero mode eq (\ref{eq:rszeromode}) satisfy the completeness relation
\begin{equation}
\psi_0^\ast (\zeta) \psi_0 (\zeta') + \int_0^\infty d \mu \; \psi^\ast ( \zeta; \mu) \psi (\zeta'; \mu) = \delta ( 
\zeta - \zeta' ).
\label{eq:rscompletion}
\end{equation}

{\em Summary of source-free modes in RS2.} 
In summary the scalar wave equation in RS2 spacetime has two kinds of
solutions in the absence of sources. The zero-mode solutions are labelled by their 
three dimensional wave-vector ${\mathbf k}$ and have the form
$g ( \zeta, {\mathbf r}; {\mathbf k} ) \exp (-i \omega t )$. The mode function $g$ is given by
\begin{equation}
g ( \zeta, {\mathbf r}; {\mathbf k} ) = \frac{1}{(2 \pi)^{3/2}} 
\frac{\sqrt{2}}{\gamma} 
\exp ( i {\mathbf k}
\cdot {\mathbf r} ).
\label{eq:zeromoders}
\end{equation}

Zero modes have the dispersion relation $\omega = k$ characteristic of waves in four spacetime dimensions. 
The second class of solutions are the Kaluza-Klein modes labelled by their three dimensional 
wave-vector ${\mathbf k}$ and their mass parameter $\mu$. The Kaluza-Klein modes have
the form $f ( \zeta, {\mathbf r}; {\mathbf k}, \mu) \exp (-i \omega t)$. The mode function $f$ is given by
\begin{equation}
f( \zeta, {\mathbf r}; {\mathbf k}, \mu) = \frac{1}{(2 \pi)^{3/2}} \zeta^{3/2} \psi( \zeta; \mu ) \exp ( i {\mathbf k}
\cdot {\mathbf r} ).
\label{eq:kkrs}
\end{equation}
$\psi( \zeta; \mu)$ is given by eq (\ref{eq:rscontinuum}). The Kaluza-Klein modes have the $\omega$-${\mathbf k}$
dispersion relation $\omega = \sqrt{ k^2 + \mu^2 }$ corresponding to the massive wave equation with mass 
parameter $\mu$. In contrast to the ADD model, which had a discrete set of Kaluza-Klein modes with quantized
mass parameters, the RS2 model has a continuum of Kaluza-Klein modes and the mass parameter
can take on any positive real value.

\indent D. {\em Static point source solution.}   
Finally let us consider the solution to a static point source confined to the brane.
The resulting field $\phi$ will also be time independent. 
We wish to show, following \cite{rs2}, 
that the field produced by such a source falls off inversely with distance at 
distances that are long compared with $\gamma^{-1}$. Hence in the RS2 model too
the extra dimension is hidden on long length scales. 

For a point source localized on the brane,
$s = \lambda \delta ({\mathbf r}) \delta ( \zeta - \gamma^{-1})$.
Taking into account that the solution $\phi$ will be time-independent, 
eq (\ref{eq:rswaveequation}) has the form 
\begin{equation}
- \nabla_{{\rm RS}}^2 \phi = \lambda \delta ({\mathbf r}) \delta (\zeta - \gamma^{-1})
\label{eq:staticrs}
\end{equation}
where
\begin{equation}
 \nabla_{{\rm RS}}^2 = 
 \left(
\frac{ \partial^2}{\partial \zeta^2} - \frac{3}{\zeta} \frac{\partial}{\partial \zeta} + 
\frac{\partial^2}{\partial x^2} + \frac{\partial^2}{\partial y^2} + \frac{\partial^2}{\partial z^2} \right). 
\label{eq:rslaplacian}
\end{equation}
In order to solve eq (\ref{eq:staticrs}) we write the solution $\phi({\mathbf r}, \zeta)$
as a superposition of zero-mode and Kaluza-Klein
mode functions. Thus
\begin{eqnarray}
\phi( \zeta, {\mathbf r}) & = & \int d {\mathbf k} \;
\tilde{\phi}_0({\mathbf k}) g (\zeta, {\mathbf r}; {\mathbf k}) \nonumber \\
& + &
\int d {\mathbf k} \int_0^\infty d \mu \; 
\tilde{\phi} ({\mathbf k}, \mu) f(\zeta, {\mathbf r}; {\mathbf k}, \mu).
\label{eq:rssuperpos}
\end{eqnarray}
The amplitudes $\tilde{\phi}_0 ({\mathbf k})$ and $\tilde{\phi} ({\mathbf k}, \mu)$ are to be determined. 

It is now helpful to note that 
\begin{eqnarray}
- \nabla^2_{{\rm RS}} \; g (\zeta, {\mathbf r}; {\mathbf k} ) & = & k^2 g,
\nonumber \\
- \nabla^2_{{\rm RS}} \; f (\zeta, {\mathbf r}; {\mathbf k}, \mu) & = & ( k^2 + \mu^2 ) f.
\label{eq:modelaplacians}
\end{eqnarray}
Eq (\ref{eq:modelaplacians}) follows because $g \exp (-i \omega t)$ and 
$f \exp (-i \omega t )$ obey the scalar wave eq (\ref{eq:rswaveequation}), 
and from the dispersion relations for the zero and Kaluza-Klein modes respectively.
Using eqs (\ref{eq:rssuperpos}) and (\ref{eq:modelaplacians}) we can write the left hand side of eq (\ref{eq:staticrs})
as
\begin{eqnarray}
- \nabla^2_{{\rm RS}} \; \phi( \zeta, {\mathbf r}) & = & \int d {\mathbf k} \;
k^2 \tilde{\phi}_0({\mathbf k}) g (\zeta, {\mathbf r}; {\mathbf k}) \nonumber \\
& + &
\int d {\mathbf k} \int_0^\infty d \mu \; 
(k^2 + \mu^2) \tilde{\phi} ({\mathbf k}, \mu) f(\zeta, {\mathbf r}; {\mathbf k}, \mu).
\nonumber \\
\label{eq:rssuperposlhs}
\end{eqnarray}
On the other hand we can obtain the right hand side of eq (\ref{eq:staticrs}) 
by substituting $\zeta' \rightarrow \gamma^{-1}$ and ${\mathbf r}' \rightarrow 0$
in the identity \footnote{The identity is merely a precise statement that the 
modes $g$ and $f$ are a complete basis. It is 
easily proved by making use of the explicit forms of $g$ and $f$ in 
eqs (\ref{eq:zeromoders}), (\ref{eq:kkrs}), as well as the completeness relation 
(\ref{eq:rscompletion}) and the standard Fourier integral
$\int d {\mathbf k} \; \exp[ i {\mathbf k} \cdot ( {\mathbf r} - {\mathbf r}') ] = (2 \pi)^3 \delta ({\mathbf r} - {\mathbf r}')$.}
\begin{eqnarray}
& & \zeta'^3 \delta (\zeta - \zeta') \delta ({\mathbf r} - {\mathbf r}') = 
\int d {\mathbf k} \; g^\ast (\zeta', {\mathbf r}'; {\mathbf k} ) g (\zeta, {\mathbf r}; {\mathbf k} )
\nonumber \\
& & +  \int d {\mathbf k} \int_0^\infty d \mu \; 
f^\ast (\zeta', {\mathbf r}'; {\mathbf k}, \mu) f (\zeta, {\mathbf r}; {\mathbf k}, \mu).
\label{eq:identity}
\end{eqnarray}
Making these substitutions in eq (\ref{eq:identity}) and equating to eq (\ref{eq:rssuperposlhs}) yields
\begin{eqnarray}
\tilde{\phi}_0 ({\mathbf k}) & = & \lambda 
\frac{\gamma^3}{k^2} g^\ast \left( \frac{1}{\gamma}, {\boldsymbol 0}; {\mathbf k} \right),
\hspace{3mm}
\nonumber \\
\tilde{\phi} ({\mathbf k}, \mu) & = & \lambda 
\frac{\gamma^3}{k^2 + \mu^2} f^\ast \left( \frac{1}{\gamma}, {\boldsymbol 0};
{\mathbf k}, \mu \right).
\label{eq:rsamplitudes}
\end{eqnarray}
. 

Substituting eq (\ref{eq:rsamplitudes}) in eq (\ref{eq:rssuperpos}) and making use of the 
explicit forms of $g$ and $f$ given in eqs (\ref{eq:zeromoders}) and (\ref{eq:kkrs}) we 
obtain
\begin{eqnarray}
\phi (\zeta, {\mathbf r}) & = & 2 \lambda \gamma \int \frac{ d {\mathbf k} }{(2 \pi)^3} 
\frac{1}{k^2} \exp ( i {\mathbf k} \cdot {\mathbf r} ) 
\nonumber \\
& + & \lambda \gamma \zeta^2 \int \frac{ d {\mathbf k}}{(2 \pi)^3} \int_0^\infty d \mu \; 
\frac{1}{k^2 + \mu^2} \exp ( i {\mathbf k} \cdot {\mathbf r} ) 
\nonumber\\
& & \times \left[ \alpha (\mu) J_2 \left( \frac{ \mu }{\gamma} \right) + \beta ( \mu ) Y_2 \left( \frac{\mu}{\gamma} \right) \right]^2. 
\label{eq:rspointalmost}
\end{eqnarray}
The first term corresponds to the zero-mode contribution the field; the second to the Kaluza-Klein modes.
Performing the
integrals over ${\mathbf k}$, which are the same as in eqs (\ref{eq:fouriercoulomb}) and (\ref{eq:yukawa}), yields
\begin{eqnarray}
\phi ( \zeta, {\mathbf r} ) & = & \frac{ \lambda \gamma }{2 \pi r} 
 +  \frac{\lambda \gamma \zeta^2}{4 \pi r} \int_0^\infty d \mu \; \exp ( - \mu r ) 
 \nonumber \\
 & & \times
\left[ \alpha (\mu) J_2 \left( \frac{ \mu }{\gamma} \right) + \beta ( \mu ) Y_2 \left( \frac{\mu}{\gamma} \right) \right]^2. 
\label{eq:rspointexact}
\end{eqnarray}
Eq (\ref{eq:rspointexact}) is the exact expression for the field due to a point source located at the brane.
It is instructive to simplify it in the limit of long distance $r \gg \gamma^{-1}$. In this limit, due to the 
decaying exponential it is acceptable to approximate the remaining terms in the integrand by their
small $\mu$ limit. The expansion is tedious by hand but Mathematica readily finds that the 
square of the term in square brackets in eq (\ref{eq:rspointexact}) is approximately equal to $\mu$ in 
the small $\mu$ limit. Making this approximation and evaluating the integral over $\mu$ we finally obtain
\begin{equation}
\phi ( \gamma^{-1}, {\mathbf r} ) \approx \frac{ \lambda \gamma }{ 2 \pi r} \left( 1 + \frac{1}{2 \gamma^2 r^2} + \ldots \right)
\label{eq:pointsourcefarrs}
\end{equation}
for $r \gg \gamma^{-1}$. We have also set $\zeta \rightarrow \gamma^{-1}$ since we are primarily interested
in the field on the brane. 
Thus as advertised at long distances a point source couples primarily to the zero mode and
produces a field that varies as $1/r$. Thus the extra dimension is hidden. The leading correction due to coupling to the
Kaluza Klein modes varies as $1/r^3$ rather than exponentially as in the ADD model.

In principle we can work out the small $r$ behavior also from eq (\ref{eq:rspointexact}). 
In practice it is simpler 
to make the following physical argument. Locally
any space or spacetime looks flat. 
The expression for the field of a point source in a flat four dimensional space can
be obtained by taking the $L \rightarrow \infty$ limit of eq (\ref{eq:addpointsolution})
which is
the same as the $r \ll L$ result, eq (\ref{eq:oneoverrsquare}).
Thus we conclude that  
\begin{equation}
\phi (\gamma^{-1}, {\mathbf r} ) \approx \frac{ \lambda}{4 \pi^2 r^2}
\label{eq:pointsourcenearrs}
\end{equation}
for $r \ll \gamma^{-1}$ in the RS model. 

Randall and Sundrum assumed that standard model
particles and fields were confined to the brane. Only gravity was assumed to
propagate through the bulk. Here we have analyzed a bulk scalar field as a 
proxy for the gravitational field. Scalar fields are simpler to analyze but have
useful parallels to gravitational fields. For example gravitational waves also
have a discrete zero mode and a continuum of Kaluza-Klein modes; indeed
the same Schr\"{o}dinger analogy and the same Bessel functions arise in both
cases. 

Suppose now that we are interested in the gravitational field of a point mass $m$ located 
on the brane. At short distances we expect its Newtonian gravitational potential will have
the five dimensional form $G_5 m/r^2$ where $G_5$ is the gravitational coupling constant.
On the other hand at long distances we expect the Newtonian potential to have the 
four dimensional form $Gm/r$ where $G$ is the familiar Newtonian constant of gravity (see fig. 89 of
ref \cite{randall} for a diagram of the field lines of a point source in RS2 spacetime).
Comparing these forms to eqs (\ref{eq:pointsourcefarrs}) and (\ref{eq:pointsourcenearrs}) we 
infer that $\lambda = 4 \pi^2 G_5 m = 2 \pi G m/ \gamma$ which implies 
\begin{equation}
G_5 = \frac{G}{2 \pi \gamma}.
\label{eq:rshierarchy}
\end{equation}
Eq (\ref{eq:rshierarchy}) agrees with eq (7) of ref \cite{rs2} and is analogous to eq (\ref{eq:hierarchysolved}) 
for the ADD model and is the key to resolving the hierarchy problem within this model. 

{\em RS1 and the Hierarchy Problem.}
The model we have analyzed thus far was introduced in \cite{rs2} and is sometimes
called the RS2 model. In their earlier paper \cite{rs1} Randall and Sundrum introduced a slightly different 
model called RS1 which provides an especially elegant resolution of the hierarchy problem. 
In this model the geometry of spacetime is still given by eq (\ref{eq:rsmetric}) but the coordinate
$w$ only extends over the range $0 \leq w \leq \ell$. In addition to the brane at $w = 0$, a second brane is
located at $w = \ell$. We shall refer to these branes as the left brane and the right brane respectively. 
In conformal co-ordinates the left brane is located at $\zeta = \gamma^{-1}$ and the right brane at
$\zeta = \gamma^{-1} \exp ( \gamma \ell )$. 
In RS1 the zero mode (\ref{eq:zeromoders}) remains a solution that satisfies the boundary condition at both branes.
The Kaluza-Klein modes become quantized: it is only for particular values of $\mu$ that the solutions
(\ref{eq:kkrs}) obey the boundary conditions at both branes. The lowest allowed value of $\mu$ is of order
$\gamma \exp ( - \gamma \ell )$ and the spacing between successive allowed values of $\mu$ is also of
this order (see problem 5 in appendix C for a derivation \cite{som}). 
In RS1 we choose $\gamma^{-1}$ and $\ell$ to both be of order the Planck
length, but if we choose the product $\gamma \ell \approx 35$-$40$, then the lowest allowed $\mu$ is sixteen orders
of magnitude smaller, and is of the weak scale. Thus a large hierarchy can emerge in RS1 due to the exponential
warp factor even though the basic parameters of the model are all of Planck scale. 
Problem 6 in Appendix C \cite{som} provides further insight into the resolution of the hierarchy problem in RS1.
It is shown there that the natural mass for a standard model particle confined to the right brane is of the
electroweak scale even though all the parameters in the RS1 model are of Planck scale. See section
\ref{sec:discussion} for further discussion of the experimental implications and tests of the RS models.

\section{DGP Model}
According to the current cosmological paradigm
about 70\% of the total energy density of the universe is in the form of a mysterious component called ``dark energy''. The first observational evidence for dark energy came from supernovae \cite{redshift}; distant supernovae appeared dimmer than expected and this could be explained by accelerated cosmic expansion. 
The DGP model was introduced as a alternative to the idea that dark energy is an actual component of the Universe. 
In the DGP model gravity becomes modified on distance scales that are long compared to $\ell$,
a parameter of the DGP model. 
On distance scales short compared to $\ell$ spacetime appears
four-dimensional and gravity obeys Newton's inverse square law. 
On distance scales long compared  to
$\ell$ the five dimensionality of spacetime is revealed; the gravitational force now falls off much
more rapidly, as the inverse cube of the distance, making  gravity less effective on cosmological scales. This could help explain the supernova observations but provides no explanation for why the parameter $\ell$ has the value
that is implied by observations.
Note that in contrast to the ADD and RS models in the DGP model the higher dimensional behavior
is manifested on long length scales rather than short.

A.{\em The spacetime interval.} In the DGP model, like in ADD and RS, there is a fourth spatial dimension, $w$.  
The spacetime interval between nearby events at in DGP is given by
\begin{equation}
ds^2 = d t^2 - dx^2 - dy^2 - dz^2 - d w^2.
\label{dgpmetric}
\end{equation}
All coordinates have the same infinite range $- \infty < t, x, y, z, w < \infty$. Thus DGP spacetime is both flat and possesses an infinite extra dimension. 
As in ADD and RS, the four dimensional section $w = 0$ is the brane; standard model particles and fields
are confined to it, while gravity propagates throughout the bulk. 
Although for simplicity we do not consider this case here, it is consistent to restrict $w$ to the
range $w > 0$ as in the Randall-Sundrum model. Indeed in the original DGP paper this restriction was
effectively imposed in the form of a boundary condition that the solutions had to be symmetric functions of
$w$.

\indent B. {\em The wave equation.} In the DGP model a scalar field that propagates in the bulk is taken to 
satisfy the wave equation
\begin{eqnarray}
& & 
\left( \frac{ \partial^2 }{\partial t^2} - \frac{\partial^2}{\partial x^2} - \frac{\partial^2}{\partial y^2} - 
\frac{\partial^2}{\partial z^2} - \frac{\partial^2}{\partial w^2}
\right) \phi 
\nonumber \\
& & 
+ \ell \delta (w) \left( 
\frac{\partial^2}{\partial t^2} - \frac{\partial^2}{\partial x^2} - \frac{\partial^2}{\partial y^2} - 
\frac{\partial^2}{\partial z^2} \right) \phi = s( {\mathbf r}, w, t).
\nonumber \\
\label{dgpwaveequation}
\end{eqnarray}
Eq. (\ref{dgpwaveequation}) has the form of the wave equation one would expect from 
the spacetime interval in Eq. (\ref{dgpmetric}) only if we set $\ell = 0$. 
The term in Eq. (\ref{dgpwaveequation}) that is proportional to $\ell$ is responsible for the
remarkable behavior of the DGP model described above. At the end of this section 
we give a physical motivation for the form of this
term by an analogy to the electrostatics of a dielectric 
sheet. The analogy shows that much as a dielectric sheet will screen a point charge, 
so also in the DGP model
gravity is screened by the brane at short distances. The screening leads to 
behavior at short distances appropriate to three spatial dimensions, 
but at long distances gravity is unscreened, yielding behavior appropriate
to four spatial dimensions. 

\indent C. {\em Source free solution}.
We seek solutions to eq (\ref{dgpwaveequation}) of the form
$\upsilon(w) \exp ( i {\mathbf k} \cdot {\mathbf r} ) \exp ( - i \omega t )$. Substituting this ansatz
into eq (\ref{dgpwaveequation}) we obtain
\begin{equation}
- \frac{1}{2} \frac{d^2}{d w^2} \upsilon - \frac{\ell \mu^2}{2} \delta (w) \upsilon = \frac{\mu^2}{2} \upsilon.
\label{eq:dgpmodequation}
\end{equation}
Here as usual we have written $\omega^2 = k^2 + \mu^2$.
Superficially this resembles the Schr\"{o}dinger equation for a non-relativistic particle moving in 
one dimension under the influence of an attractive delta function potential. The analogy is not
perfect because the strength of the delta potential depends on $\frac{1}{2} \mu^2$ which is analogous
to the ``energy''. Nonetheless we can draw upon experience with
delta potentials in quantum mechanics in solving eq (\ref{eq:dgpmodequation}). We expect a solution 
of the scattering form
\begin{eqnarray}
\upsilon_+(w; \mu) & = & \frac{1}{\sqrt{2 \pi}} \exp (i \mu w) + \frac{r(\mu)}{\sqrt{2 \pi}} 
\exp( - i \mu w) \hspace{2mm} {\rm for} \hspace{2mm}
w < 0
\nonumber \\
& = & \frac{t(\mu)}{\sqrt{2 \pi}} \exp ( i \mu w ) \hspace{4mm} {\rm for} \hspace{2mm} w > 0. 
\label{eq:scattering}
\end{eqnarray}
There should be a second solution, $\upsilon_-(w; \mu)$,
corresponding to a plane wave incident along the negative $w$ axis. 
The detailed analysis of these solutions is relegated to the problems. Here we note that the
``transmission'' coefficient is found to be
\begin{equation}
| t ( \mu ) |^2 = \frac{1}{1 + (\mu \ell/2)^2}.
\label{eq:lorentzian}
\end{equation}
There is a peak in the transmission at $\mu = 0$. The peak has a Lorentzian form and a width of $1/\ell$.
Remarkably, in the DGP model when there is no source, there is no zero mode. But there 
is a continuum of massive Kaluza-Klein modes and the mass parameter $\mu$ can take any positive real value. 
Although there is no discrete zero-mode bound to the brane, 
there is a resonance in the Kaluza-Klein modes at $\mu =0 $ \footnote{As discussed in
introductory textbooks, a resonance is an
almost-bound state in quantum mechanics. As a concrete example consider a non-relativistic particle
moving in one dimension in a double barrier potential. If the barrier walls were infinitely high there would
be bound states between the barriers. For finite barriers these states are no longer bound and become
resonances. They acquire a finite
lifetime as the particle can escape by tunneling. If we consider scattering states in which a particle is incident on 
the double barrier potential from outside 
we find sharp Lorentzian peaks in the transmission coefficient as a function of the energy. The energy of the
peaks corresponds roughly to the energy of the resonance and the width of the peak to the lifetime.}. 

\indent D {\em Static point source solution.} We now consider the field produced by a static point source localized on the brane. 
Thus the source is $s \rightarrow \lambda \delta ({\mathbf r}) \delta (w)$. The resulting field will also be
static and we are primarily interested in the field on the brane, $\phi ({\mathbf r}, w \rightarrow 0)$. 

We obtain the exact expression for the point source profile from straightforward Fourier analysis  of Eq. (\ref{dgpwaveequation}); see  Appendix C, Problem 4 \cite{som}. 
The result is 
\begin{equation}
\phi ({\mathbf r}, 0) = \frac{\lambda}{4 \pi^2 r} \int_0^\infty d k \; \frac{ \sin (k r) }{\left( 1 + \frac{1}{2} \ell k \right)}.
\label{eq:radialintegral}
\end{equation}
At short distances $(r \ll \ell$)
\begin{equation}
\phi( {\mathbf r}, 0) \approx \frac{1}{4 \pi} 
\frac{\lambda}{\ell} \frac{1}{r}
\label{eq:dgpshort}
\end{equation}
whereas at long distances $(r \gg \ell)$
\begin{equation}
\phi({\mathbf r}, 0) \approx \frac{\lambda}{4 \pi^2} 
\frac{1}{r^2}.
\label{eq:dgplong}
\end{equation}

Intuitively we can understand these results as follows. Evidently $\phi({\mathbf r}, 0)$
must depend on $r, \ell$ and $\lambda$. On dimensional grounds we expect $\phi({\mathbf r}, 0) = (\lambda/r^2) 
f(\ell/r)$ where $f$ is an unknown function. This form suggests that to obtain the $\ell \ll r$ limit it is sufficient to
consider the $\ell \rightarrow 0$ limit of Eq (\ref{dgpwaveequation}). In that limit
eq (\ref{dgpwaveequation}) reduces to the electrostatics of a point source 
in four space dimensions. Hence $\phi$ has the same form as in the small $r$ limits of the 
ADD and RS models; compare eqs (\ref{eq:oneoverrsquare}), (\ref{eq:pointsourcenearrs}) 
and (\ref{eq:dgplong}). 
Conversely it is plausible that the $\ell \gg r$ limit can be derived by neglecting the first term on the left hand
side of Eq (\ref{dgpwaveequation}). In that case we are simply solving Poisson's equation in ordinary
three dimensional space for the potential of a point charge $\lambda/\ell$ thereby justifying 
Eq (\ref{eq:dgpshort}). Remarkably we obtain behavior appropriate to four spacetime dimensions at short distances
and to five spacetime dimensions at long distances.
In contrast to the ADD and RS models, here the extra dimension reveals itself at long distances
and remains hidden at short.

{\em Dielectric sheet analogy.} 
Finally we motivate the form of the wave Eq. (\ref{dgpwaveequation}). 
To this end we proceed by analogy to the electrostatics
of a two dimensional dielectric sheet with a point charge
$q$ place at the origin. As a result the bound charges in the
dielectric sheet are polarized. The polarization vector lies
in the plane of the sheet and is given by
\begin{equation}
P_x = - \ell \frac{\partial}{\partial x} \phi \; \delta (z) \hspace{3mm} {\rm and} \hspace{3mm}
P_y = - \ell \frac{\partial}{\partial y} \phi \; \delta (z).
\label{eq:polarization}
\end{equation}
Here $\ell$ denotes the polarizability of the sheet, which is assumed to lie in the $x$-$y$ plane,
and $\phi$ is the electrostatic scalar potential.
The resulting bound charge density is then given by the negative divergence of the polarization. Thus
\begin{equation}
\rho_{{\rm pol}} = \ell \left( \frac{\partial^2}{\partial x^2} \phi + \frac{\partial^2}{\partial y^2} \phi \right) \delta (z).
\label{eq:sheetdensity}
\end{equation}
By Gauss's law $\nabla \cdot {\mathbf E} = q \delta ({\mathbf r}) + \rho_{{\rm pol}}$. Thus Poisson's
equation takes the form
\begin{equation}
- \left[ 
\frac{\partial^2}{\partial x^2} + \frac{\partial^2}{\partial y^2} + \frac{\partial^2}{\partial z^2} \right] \phi
- \ell \delta (z) \left[
\frac{\partial^2}{\partial x^2} + \frac{\partial^2}{\partial y^2} \right] \phi = q \delta ({\mathbf r}).
\label{poissondgp}
\end{equation}
Eq (\ref{poissondgp}) is easily solved using Fourier analysis. The result is that at short distances from the source the
field is screened and appears to be that of a line of charge but at long distances it is unscreened and 
reduces to the field of a point charge. 
Note that Eq. (\ref{dgpwaveequation}) is a natural generalization of 
Eq.(\ref{poissondgp}) that incorporates dynamics by including time derivatives and that also incorporates an extra dimension of space.
 
\vspace{-2.5mm}

\section{Discussion} 
\label{sec:discussion}
Table 1 summarizes the main properties of the ADD, RS and DGP models discussed here.
Having demonstrated the utility of scalar fields for providing
insights into the physics of these models we now briefly
discuss the motivation for these models and their experimental
tests. A valuable resource on experiments is ref \cite{particledata}
which provides a lucid and authoritative review that is regularly updated.

The standard model of particle physics is believed to be a low energy approximation to a more
fundamental theory that remains to be discovered. Grand unification is a longstanding and ambitious program to 
develop a unified theory of the strong and electroweak forces \cite{wilczek}. Within this paradigm
the strong and electroweak forces are unified at an energy scale $E_{{\rm GUT}} \sim 10^{25} {\rm eV} \sim
10^{-3} E_P$. This energy scale is beyond the reach of any collider of the near future. Physics on the
grand unification scale is only accessible to precision experiments such as proton decay or via
cosmology. Crucially with the grand unification paradigm $E_{{\rm GUT}}$ and $E_P$ are well
separated and gravity is a separate problem not addressed by grand unification.

Extra dimensional models emerged more recently as an alternative to grand unification and
represent a very different view of the world.
As discussed above, according to the ADD model, the fundamental scale for the unification of
all forces, including gravity, is the weak scale, not $E_{{\rm GUT}}$ or $E_{{\rm P}}$.
According to the RS model too, although the fundamental
scale for gravity is the Planck scale, quantum gravity effects should be 
accessible to colliders in the form of TeV scale Kaluza-Klein
particles as we discuss below. 

Precision tests of Newton's inverse square law with torsion pendulums represent an important
class of experimental tests of the ADD model \cite{adelberger}. Both RS and ADD predict that the gravitational
force will depart from Newton's inverse square law at small distances; however for the RS models this happens
at distances that are too small to be observable. For the ADD model such experiments easily preclude the
existence of a single extra dimension of order $10^{13}$m. For the case of two extra dimensions the best
upper bound is $L < 60 \mu$m \cite{eotwash}. 

Another important class of experiments aim to detect the Kaluza-Klein modes either directly or
indirectly. Taking into account quantum mechanics, in both ADD and RS models the zero mode corresponds
to a massless particle---the graviton. Each Kaluza-Klein mode can be thought of as a separate particle with
a mass equal to $\mu \hbar/c$ where $\mu$ is the mass parameter of the mode. Thus both models
predict an infinite tower of new particles. The mass of the lightest Kaluza-Klein particle in the ADD model 
would be as low as $10^{-4}$ eV assuming two extra dimensions of millimeter scale (since $\mu \sim 1/L$ 
for the ADD model). This is also the scale of the separation between in mass between successive 
Kaluza-Klein particles. In RS1 the lightest Kaluza-Klein particle has a much larger mass of order
1 TeV (since $\mu \sim \gamma e^{-\gamma \ell}$) and that is also the order of the spacing between
successive Kaluza-Klein particle masses. Furthermore, though this is not evident from our presentation,
the Kaluza-Klein particles in RS1 interact strongly with standard model particles on the TeV scale,
whereas the interaction with the corresponding ADD particles is only of gravitational strength. Thus the
Kaluza-Klein modes in RS1 would be manifested in collider experiments as well-defined particles with
masses on the TeV scale. The Kaluza-Klein modes of the ADD model would not be
individually observable, but because they are so numerous their production would be 
manifested as missing energy in collisions \cite{particledata}. Experiments at LHC have put an 
upper bound of 50-100 fm for ADD models with between four to six extra dimensions \cite{atlas}.  

Astrophysical observations are also sensitive to the production of Kaluza-Klein particles in the
ADD model. For example, if the production becomes too copious it would put supernovae models in conflict
with data \cite{particledata, hannestad}. Observations of neutron stars by the Fermi gamma ray observatory
provide another astrophysical constraint \cite{ajello}. For the case of two extra dimensions the best current astrophysical
constraints are of order nanometers. Thus although indirect these constraints can be quite stringent. 

In contrast to the ADD and RS models the DGP model is concerned with gravity on the cosmological
scale. The DGP model predicts modifications of gravity on length scales that are long compared to the 
Hubble scale, 
$\ell \sim 10^9$ light years. The natural tests of the DGP model therefore come from cosmology. 
The DGP model is now disfavored by observations of the accelerated expansion of the Universe
\cite{hu} but it has many descendants that remain viable \cite{claudia}.

Finally we note that the beyond the immediate application to particle physics
the RS model has deep connections to string theory. An important development
in string theory is the AdS-CFT correspondence that demonstrates a deep relationship
between theories defined in the bulk of AdS spacetime and on its boundary \cite{zwiebach}. 
This correspondence further illuminates the RS model. Many interesting
applications of the AdS-CFT correspondence are in condensed matter physics \cite{subir}.
Thus extra dimensional models sit at a vibrant intersection of theory and 
experiment and different subfields of physics.

{\em Acknowledgements.} We acknowledge discussions with Walter Lambrecht, Raman Sundrum 
and Andrew Tolley. We also are very grateful to the referees for many insightful comments and suggestions.

\begin{table*}[h!]
\centering
\rotatebox{0}{
\resizebox{\textwidth}{!}{
\begin{tabular}{|  1 | 3 | 4 |  2 | 2 | 5 | }
\hline
\;{\bf Model}&
\multicolumn{2}{ c| }{\bf Source-Free Solutions:} &\;{\bf Point Source Profile}&\;{\bf Comments}\\ \cline{2-3}

 &Mode Types &  Particle Interpretation \newline \& Dispersion Relation &  &  \\ \hline
\;{\bf Massless Scalar Field}\newline \;(3+1)&
\; Plane waves \newline & \;Massless with $\omega=k$  &\;$\;\;\;\;\;\;\;\;\phi \propto 1/r$\newline \newline (Eq.4)&\;Ordinary spacetime of special relativity.\\ \hline
 \;{\bf Massive Scalar Field}\newline\;(3+1) &
 \;Plane waves\newline & \;Massive with $\omega = \sqrt{k^2+\mu^2}$ &\;$\;\;\;\;\;\;\phi \propto e^{-\mu r}/r$\newline \newline (Eq.6)& \;Reduces to massless for $\mu \rightarrow 0.$ \\ \hline 
\; $\;\;\;${\bf ADD}\newline$(d+1)$&
\;Zero mode \rule{8.7cm}{.08mm}  \newline \newline KK modes: discrete & \;Massless with $\omega=k$\newline  \newline \newline Massive with $\omega = \sqrt{k^2+\kappa_n^2}$ &\[
    \phi \propto 
\begin{cases}
    1/r& \text{for $ r \gg L$} \\
    1/r^2& \text{for $r \ll L$}
\end{cases}
\](Eq. 16,17)&\;Solves hierarchy problem via $G_5=GL/\pi$ and $E_{EW} = E_{P5}$ 
for single extra dimension of size $L$.\\ \hline
\; $\;\;\;\;${\bf RS}\newline \;(4+1)&
\;Zero mode \rule{8.7cm}{.08mm} \newline \newline KK modes: \newline discrete $\;$(RS1)\newline continuum (RS2) & \;Massless with $\omega=k$\newline \newline \newline   Massive with $\omega = \sqrt{k^2+\mu^2}$ &\[
    \phi \propto 
\begin{cases}
    1/r& \text{for $r \gg \gamma^{-1}$} \\
    1/r^2& \text{for $r \ll \gamma^{-1}$}
\end{cases}
\](Eq. 40,41)&\;RS1 solves hierarchy problem via $E_{EW} = E_P e^{-\gamma \ell}$ \newline
for extra dimension of size $\ell$ and
curvature $\gamma$ \\ \hline
\; $\;\;\;${\bf DGP}\newline(4+1)&
\;No zero modes but a resonance at $\mu=0$ \rule{8.7cm}{.08mm}  \newline KK modes: continuum &\;width of peak $\sim 1/\ell$ \newline \newline \newline \newline Massive with $\omega = \sqrt{k^2+\mu^2}$ &\[
    \phi \propto 
\begin{cases}
    1/r& \text{for $r \ll \ell$} \\
    1/r^2& \text{for $r \gg \ell$}
\end{cases}
\](Eq. 49,50)&\;Solves dark energy problem for screening length \newline $\ell \sim$ Hubble length.\\ \hline
\end{tabular}}}
\caption{Comparison and summary of all the models studied in this paper. For the ADD model the space dimensionality
$d = 4, 5, 6, \ldots$ In the table, for illustrative purposes, we take $d=4$.}
\end{table*}

\newpage

 \appendix

\section{Differential Geometry}

{\em Curved space.} Let us start with ordinary three dimensional Euclidean space. Adopting cartesian 
co-ordinates the distance between neighboring points is given by the Pythagorean formula
$d s^2 = d x^2 + dy^2 + dz^2$. Suppose
however we use spherical polar co-ordinates $(r, \theta, \varphi)$ to label points in the same space. 
The distance between neighboring points in spherical polar co-ordinates is given by
\begin{equation}
d s^2 = d r^2 + r^2 d \theta^2 + r^2 \sin^2 \theta d \varphi^2.
\label{eq:polarmetric}
\end{equation}
Eq (\ref{eq:polarmetric}) follows readily from the transformation that given Cartesian co-ordinates
in terms of polar; namely, $z = r \cos \theta$, $x = r \sin \theta \cos \varphi$ and $y = r \sin \theta \sin \varphi$. 
Regardless of the co-ordinates we use, the geometry of the space is fully specified by the formula for
distance between neighboring points. 

Thus far we are only describing the same flat space in different co-ordinates. 
Now let us discussed curved spaces. The simplest example is 
the surface of a sphere of radius $R$ that is located at the origin of the Cartesian co-ordinate
system. Points on the surface of the sphere can be labelled by the co-latitude $\theta$ and the longitude
$\varphi$. The distance between neighboring points on the surface of a sphere is given by 
\begin{equation}
d s^2 = R^2 d \theta^2 + R^2 \sin^2 \theta \; d \varphi^2.
\label{eq:spheremetric}
\end{equation} 
The co-ordinates are restricted to lie in the range $0 \leq \theta \leq \pi$ and $0 \leq \varphi < 2 \pi$. 
Although it is helpful to picture the sphere as a surface in three dimensional Euclidean space mathematically
the space is fully defined by eq (\ref{eq:spheremetric}) together with a statement about the allowed range of
the co-ordinates. 

Another example of a curved space is the two dimensional hyperbolic space. Points in this space
are labelled by the co-ordinates $(\theta, \varphi)$ which have the ranges $0 \leq \theta < \infty$ 
and $0 \leq \varphi < 2 \pi$. The distance between neighboring points on the surface of hyperbolic
space is given by
\begin{equation}
d s^2 = R^2 d \theta^2 + R^2 \sinh^2 \theta \; d \varphi^2.
\label{eq:hyperbolic}
\end{equation}
Hyperbolic space cannot be regarded as a surface in ordinary three dimensional space but it can be 
embedded in a different flat three dimensional space in which the co-ordinates $(x, y, z)$ have the
usual range $ - \infty < x, y, z < \infty$ but the distance between neighboring points is given by
\begin{equation}
d s^2 = - dz^2 + dx^2 + dy^2 
\label{eq:negativedistance}
\end{equation}
This space is flat but it has the peculiar feature that the square of the distance between 
nearby points can be positive, negative or zero. Mathematically this space is said to have
an indefinite metric. The hyperbolic space is the two dimensional surface that satisfies the equation
\begin{equation}
z^2 - x^2 - y^2 = R^2
\label{eq:hyperboloid}
\end{equation}
together with the condition $z > 0$. It is easy to see that the constraint eq (\ref{eq:hyperboloid}) is
automatically satisfied if we write 
\begin{equation}
z = R \cosh \theta, x = R \sinh \theta \cos \varphi,  y = R \sinh \theta \sin \varphi.
\label{eq:hyperbolicparameters}
\end{equation}
Furthermore the entire hyperboloid will be covered by this
parametrization if $(\theta, \varphi)$ are allowed the range noted above.
It is now a simple exercise to see that the distance formula eq (\ref{eq:hyperbolic}) follows
when eq (\ref{eq:hyperbolicparameters}) is substituted into eq (\ref{eq:negativedistance}). 

{\em The Laplacian.} Let us now return to ordinary three dimensional space inhabited by a scalar
field such as the scalar potential, $\phi (x,y,z)$. It is a familiar result of elementary calculus 
that the gradient of a scalar
field is a vector; and that the divergence of the gradient, which is the Laplacian, is a scalar. 
In symbols, $\nabla \cdot \nabla \phi = \nabla^2 \phi$ is a scalar. In Cartesian co-ordinates
the Laplacian may be written as 
\begin{equation}
\nabla^2 \phi = 
\frac{\partial^2}{\partial x^2} \phi + \frac{\partial^2}{\partial y^2} \phi + \frac{\partial^2}{\partial z^2} \phi.
\label{eq:cartesianlaplacian}
\end{equation}
It is helpful sometimes to work in a different co-ordinate system such as polar or cylindrical co-ordinates
and therefore it is useful to be able to express the Laplacian directly in other co-ordinate systems. 
Suppose we transform to a system of co-ordinates $(\xi, \eta, \zeta)$ in which the distance between
neighboring points is given by
\begin{equation}
d s^2 = h_\xi^2 d \xi^2 + h_\eta^2 d \eta^2 + h_\zeta^2 d \zeta^2.
\label{eq:scalefactors}
\end{equation}
The distance formula eq (\ref{eq:scalefactors}) can easily be derived from the Pythagorean formula
$ds^2 = dx^2 + dy^2 + d z^2$ given $(\xi, \eta, \zeta)$ as functions of $(x,y,z)$. In general the formula
might involved cross terms like $d \xi d \eta, d \eta d \zeta$ and $d \zeta d \xi$ but here we will focus
only on orthogonal co-ordinates in which such terms are absent. Most co-ordinate systems of interest
in mathematical physics are orthogonal. The factors $h_\xi, h_\eta$ and $h_\zeta$ are called scale
factors and their product $h = h_\xi h_\eta h_\zeta$ is called the invariant measure.
(For  readers 
who have studied general relativity we note, parenthetically, that $h$ is commonly denoted $\sqrt{|g|}$ where $g$ is the
determinant of the metric tensor). For example
for spherical polar co-ordinates the scale factors are $h_r = 1, h_\theta = r, h_\varphi = r \sin \theta$
and the invariant measure $h = r^2 \sin \theta$. Now it is proved in books on electromagnetism and
mathematical physics that the Laplacian in the co-ordinates $(\xi, \eta, \zeta)$ is given by
\begin{equation}
\frac{1}{h} \frac{\partial}{\partial \xi} \left( \frac{h}{h_\xi^2} \frac{\partial}{\partial \xi} \phi \right)
+ \frac{1}{h} \frac{\partial}{\partial \eta} \left( \frac{h}{h_\eta^2} \frac{\partial}{\partial \eta} \phi \right)
+ \frac{1}{h} \frac{\partial}{\partial \zeta} \left( \frac{h}{h_\zeta^2} \frac{\partial}{\partial \zeta} \phi \right)
\label{eq:polarlaplacian}
\end{equation}
The generalization of this result to more or fewer dimensions should be obvious. It is a good exercise 
for the reader to use eq (\ref{eq:polarlaplacian})
construct the Laplacian in familiar cases like spherical polar and cylindrical co-ordinates. 

Thus far we have been discussing the mundane subject of writing the Laplacian in 
different co-ordinate systems in ordinary flat three dimensional space. Remarkably it turns
out that in a curved space in which the distance formula has the form eq (\ref{eq:scalefactors})
it is still true that the Laplacian is given by eq (\ref{eq:polarlaplacian}). This is intuitively 
plausible and gives us a simple method to construct the Laplacian in curved space without
going through the full machinery of differential geometry \footnote{A full course in differential geometry
derives the expression for a Laplacian by first generalizing the notions of scalars, vectors and
tensors to curved spaces, developing the notion of a covariant derivative, 
and then defining the Laplacian as the scalar obtained by contracting
a second covariant derivative of a scalar field.}. 
It is a good exercise for the reader to
use eq (\ref{eq:polarlaplacian}) to write an expression for the Laplacian on the surface of a 
sphere and in hyperbolic space. 

{\em Curved space-time.} The central dogma of special relativity is that an inertial observer using
Cartesian co-ordinates can label events by the co-ordinates $(t, x, y, z)$ which lie in the range
$-\infty < t, x, y, z < \infty$ and that the interval between two nearby events is given by eq (1) 
Suppose we transform to a system of co-ordinates $(\tau, \xi, \eta, \zeta)$ in which the space-time interval 
between neighboring events is given by 
\begin{equation}
d s^2 = h_\tau^2 d \tau^2 - h_\xi^2 d \xi^2 - h_\eta^2 d \eta^2 - h_\zeta^2 d \zeta^2.
\label{eq:specialscalefactors}
\end{equation}
The interval eq (\ref{eq:specialscalefactors}) can easily be derived from eq (1) 
given $(\tau, \xi, \eta, \zeta)$ as a function of $(t, x, y, z)$. In general the formula might involve
cross terms like $d \tau d \xi$ or $d \xi d \eta$ but here we will focus only on orthogonal co-ordinates
in which such terms are absent. The factors $h_\tau, h_\xi, h_\eta$ and $h_\zeta$ are called
scale factors and their product $h = h_\tau h_\xi h_\eta h_\zeta$ is called the invariant measure. 
Thus far we are dealing with the Minkowski space-time of special relativity. Although eq (\ref{eq:specialscalefactors})
looks complicated we know that the underlying space-time is flat and that 
we can always transform back to Cartesian co-ordinate $(t, x, y, z)$ in which
the interval has the simple form eq (1). 
By contrast a curved space-time is one
in which the space-time interval might have a form like eq (\ref{eq:specialscalefactors}) but where
it is impossible to find an alternative set of co-ordinates in which the interval globally has the simple flat
form given in eq (1).

A concrete example of a curved space-time is the Friedman-Walker-Robertson space-time that is believed
to describe our expanding universe. It has the space-time interval
\begin{equation}
ds^2 = d t^2 - a^2(t) [ d x^2 + d y^2 + d z^2 ].
\label{eq:frw}
\end{equation}
Here $a(t)$ is called the scale factor and it grows with time reflecting the expansion of the Universe. 
This space-time is called flat FRW. Notwithstanding
that name it is undeniably a curved space-time. There is no change of co-ordinates which will bring the space-time 
interval (\ref{eq:frw}) to
the form eq (1). 

A second example of a curved space-time is the celebrated Anti-de-Sitter (AdS) space-time. 
Like the hyperbolic space discussed above, AdS space-time is best understood by embedding
it in a space of higher dimensionality. We start with a six dimensional flat space with the interval
\begin{equation}
ds^2 = du^2 + dv^2 - d \alpha^2 - d \beta^2 - d \eta^2 - d \xi^2.
\label{eq:adsembed}
\end{equation}
The co-ordinates have the usual range $- \infty < u, v, \alpha, \beta, \eta, \xi < \infty$. 
AdS is the five dimensional surface defined by the constraint
\begin{equation}
u^2 + v^2 - \alpha^2 - \beta^2 - \eta^2 - \xi^2 = \frac{1}{\gamma^2}.
\label{eq:adssection}
\end{equation}
We now put down new co-ordinates $(\zeta, t, x, y, z)$ on the half of AdS space-time 
that satisfies $v + \xi > 0$. The new co-ordinates are related to the old via
\begin{equation}
v + \xi = \frac{1}{\gamma \zeta}, \; 
u = \frac{t}{\gamma \zeta}, \;
\alpha = \frac{x}{\gamma \zeta}, \;
\beta = \frac{y}{\gamma \zeta}, \;
\eta = \frac{z}{\gamma \zeta}.
\label{eq:newads}
\end{equation}
The constraint eq (\ref{eq:adssection}) is automatically satisfied provided we take
\begin{equation}
v - \xi = \frac{1}{\gamma \zeta} \left( \zeta^2 + x^2 + y^2 + z^2 - t^2 \right).
\label{eq:vminusxi}
\end{equation}
The new co-ordinates have the range $\zeta \geq 0$ and $- \infty < t, x, y, z, < \infty$. 
Making use of eq (\ref{eq:newads}) and eq (\ref{eq:adsembed}) it follows after
some algebra that the interval on AdS space-time in terms of the new co-ordinate is
given by
\begin{equation}
ds^2 = \frac{1}{\gamma^2 \zeta^2} \left( d t^2 - d \zeta^2 - dx^2 - dy^2 - dz^2 \right),
\label{eq:adsconformal}
\end{equation}
exactly the form for Randall-Sundrum space-time in conformal co-ordinates. 

{\em Scalar waves in curved space-time.} A scalar wave in Minkowski space-time obeys the wave eq (2). 
The operator $\Box^2$ defined in eq (2) 
is called the 
d'Alembertian and is the space-time analog of a Laplacian. If $\phi$ is a scalar then so is
$\Box^2 \phi$ (see for example, ref [15],
vol II, chapter 25, section 3). 
Suppose we now transform to a system of co-ordinates $(\tau, \xi, \eta, \zeta)$ in which
the space-time interval between neighboring events is given by eq (\ref{eq:specialscalefactors}). 
Evidently in this co-ordinate system the d'Alembertian operator will have the form
\begin{eqnarray}
\Box^2 \phi & = & \frac{1}{h} \frac{\partial}{\partial \tau} \left( \frac{h}{h_\tau^2} \frac{\partial \phi}{\partial \tau} \right) - 
\frac{1}{h} \frac{\partial}{\partial \xi} \left( \frac{h}{h_\xi^2} \frac{\partial \phi}{\partial \xi} \right) 
\nonumber \\
& & - \frac{1}{h} \frac{\partial}{\partial \eta} \left( \frac{h}{h_\eta^2} \frac{\partial \phi}{\partial \eta} \right) - 
\frac{1}{h} \frac{\partial}{\partial \zeta} \left( \frac{h}{h_\zeta^2} \frac{\partial \phi}{\partial \zeta} \right) 
\label{eq:dalembertian}
\end{eqnarray}
Thus far we are discussing the mundane subject of writing the d'Alembertian in different co-ordinate
systems in ordinary flat Minkowski space-time. Remarkably it turns out that in a curved space-time 
in which the space-time interval has the form eq (\ref{eq:specialscalefactors}), it is still true that the
d'Alembertian is given by eq (\ref{eq:dalembertian}). This is intuitively plausible and it gives us a simple
method to write down the wave equation in curved space-times without going through the full machinery
of differential geometry. Although we have written our results specifically for a four dimensional space-time
the generalization to higher or lower dimensions is self-evident.

\section{One dimensional quantum mechanics}

{\em Normalization.} We begin by recalling some useful results from undergraduate quantum mechanics.
For derivations readers should consult their favorite textbook. 
Consider a free particle in one dimension. In a state of definite momentum 
the state of the particle is described by the wave function 
\begin{equation}
\psi(x; k) = \frac{1}{\sqrt{2 \pi}} \exp (i k x).
\label{eq:planewaveappendix}
\end{equation}
The pre-factor has been chosen to ensure the normalization
\begin{equation}
\int_{-\infty}^\infty d x \; \psi^\ast(x; p) \psi (x; k) = \delta (p - k).
\label{eq:planewavenorm}
\end{equation}
Now suppose that there is a potential barrier $V(x)$ that is localized near the origin. 
Sufficiently far from the origin the particle is free and we expect to find scattering solutions of the form
\begin{eqnarray}
\psi (x; k) & = & \frac{1}{\sqrt{2 \pi}} \exp ( i k x ) + \frac{r}{\sqrt{2 \pi}} \exp ( - i k x )
\nonumber \\
& & 
\hspace{24.5mm} {\rm for} \hspace{3mm} x \rightarrow - \infty
\nonumber \\
& = & \frac{t}{\sqrt{2 \pi}} \exp (i k x) \hspace{3mm} {\rm for} \hspace{3mm} x \rightarrow \infty
\label{eq:scatteringsolution}
\end{eqnarray}
Here the reflection coefficient $r$ and the transmission coefficient $t$ might depend on the wave-vector
$k$ and satisfy the unitarity condition $|r|^2 + |t|^2 = 1$. The scattering 
wave functions still satisfy the normalization condition eq (\ref{eq:planewavenorm}). 
The solution above corresponds to the scattering of a particle that is incident on the barrier from
the left. An analogous solution may be written down for particles incident from the right but will not
be needed here. Now suppose that the potential $V \rightarrow \infty$ as $x \rightarrow \infty$ or 
else that the particle encounters an impenetrable barrier such as a hard wall at the origin. In that
case the transmission coefficient $t \rightarrow 0$ and the reflection coefficient has magnitude 
unity and may be written as $r = \exp ( i 2 \delta)$. This defines the scattering phase shift $\delta$
which may depend on $k$. For this circumstance the scattering solution has the form
\begin{equation}
\psi (x; k) = \sqrt{ \frac{2}{\pi} } \cos ( k x - \delta ) \hspace{3mm} {\rm for} \hspace{3mm} x \rightarrow - \infty. 
\label{eq:scatteringbarrier}
\end{equation}
This form is obtained by substituting $r \rightarrow \exp ( i 2 \delta )$ in eq (\ref{eq:scatteringsolution})
and multiplying the solution by a factor of $\exp ( - i \delta )$. The scattering solutions eq (\ref{eq:scatteringbarrier})
still satisfy the normalization condition eq (\ref{eq:planewavenorm}). 

{\em Bessel solution.}
We turn now to the solution to eq (23) 
for $\mu > 0$. It is convenient to transform
to the dependent  variable $\varphi$ defined by $\psi = \zeta^{1/2} \varphi$ and to change to the 
independent variable $\xi = \mu \zeta$. In terms of these variables eq (23)
is revealed to be Bessel's equation of the second order
\begin{equation}
\frac{d^2 \varphi}{d \xi^2} + \frac{1}{\xi} \frac{d \varphi}{d \xi} + \left(1 - \frac{4}{\xi^2} \right) \varphi = 0
\label{eq:varphi}
\end{equation}
with independent solutions $J_2 (\xi)$ and $Y_2 (\xi)$. Hence the general solution to eq (23) 
has the form given in eq (25) 
with the coefficients $\alpha$ and $\beta$ at this stage arbitrary.
Imposing the boundary condition $\psi = - \frac{2}{3} \gamma^{-1} \psi'$ 
on $\psi$ at $\zeta = \gamma^{-1}$ determines the ratio $\alpha/\beta$,
\begin{equation}
\frac{\alpha}{\beta} = - Y_1 \left( \frac{\mu}{\gamma} \right) / J_1 \left( \frac{\mu}{\gamma} \right),
\label{eq:alphabetaratio}
\end{equation}
and leads to eq (26).
Here we have used the Bessel function recursion 
$x Z_1(x) = 2 Z_2 (x) + x Z_2'(x)$ where $Z$ denotes either $J$ or $Y$. 
Finally let us write 
\begin{equation}
\alpha = \sqrt{ \alpha^2 + \beta^2 } \cos \Delta \hspace{3mm} {\rm and} \hspace{3mm}
\beta = \sqrt{ \alpha^2 + \beta^2 } \sin \Delta
\label{eq:besquare}
\end{equation}
which defines $\Delta$. Inserting this form into the solution eq (25)
and making use of the
large argument asymptotics of the Bessel and Neumann functions,
\begin{equation}
J_2 (x) \approx \sqrt{ \frac{2}{\pi x} } \cos \left( x - \frac{5}{4} \pi \right), 
Y_2 (x) \approx \sqrt{ \frac{2}{\pi x} } \sin \left( x - \frac{5}{4} \pi \right),
\label{eq:besselasymptotics}
\end{equation}
we obtain 
\begin{equation}
\psi ( \zeta ) \approx
\sqrt{ \frac{ 2 (\alpha^2 + \beta^2) } { \pi \mu } } \cos \left ( \mu \zeta - \frac{5}{4} \pi - \Delta \right)
\label{eq:asymptotics}
\end{equation}
for $\mu \zeta \gg 1$. 
Comparing eq (\ref{eq:asymptotics}) to eq (\ref{eq:scatteringbarrier}) we see that our solutions 
will have the desired normalization, eq (27), 
if we choose $\alpha^2 + \beta^2 = \mu$. 
This completes the derivation of the continuum solutions to eq (23). 

For reference we note that $A$ defined in eq (26) of the paper is given by
\begin{equation}
A = \sqrt{\mu}/\sqrt{ [ Y_1 (\mu/\gamma) ]^2 +  [ J_1 (\mu/\gamma) ]^2 } 
\label{eq:ay}
\end{equation}
and therefore has the small $\mu$ asymptotic behavior 
$A \approx (\pi/2 \gamma) \mu^{3/2}$. Making use of eq (26) and eq (\ref{eq:ay}) we can also 
deduce that $ \alpha (\mu) \approx - \sqrt{\mu}$ and $\beta (\mu) \approx - (\pi/4 \gamma^2) \mu^{5/2}$
for small $\mu$. 

\section{Problems.} 

{\em Problem 1.} What data would you invoke in order to rule out the ADD model with a single extra dimension?
({\em Answer:} $10^{13}$ m is comparable to the size of Pluto's orbit. An extra dimension of this size would lead
to violations of Newton's inverse square law of gravity throughout the solar system. Tycho Brahe's data, and very
likely even Sumerian tablets, would be sufficient to rule out such violations.)

{\em Problem 2.} {\em The ADD hierarchy problem.} (a) An alternative way to describe the hierarchy problem is
in terms of length scales. (i) Construct a length scale out of the three fundamental constants $c, G$ and $\hbar$.
This is the Planck length, $\ell_{{\rm P}}$. (ii) Denote the electroweak scale $\eta = 1$ TeV. From $\eta, \hbar$
and $c$ construct the length scale $\ell_{{\rm ew}}$ that corresponds to electroweak physics. The conventional 
hierarchy problem is then the observation that $\ell_{{\rm P}} \ll \ell_{{\rm ew}}$. (b) The ADD model solves
this hierarchy problem by introducing extra dimensions of characteristic scale $L$. Compare $L$ to 
$\ell_{{\rm ew}}$ and comment. 

{\em Answer 2.} (a) $\ell_{{\rm P}} = \sqrt{ \hbar G/c^3} \sim 10^{-35}$ m. $\ell_{{\rm ew}} = \hbar c/\eta \sim 10^{-19}$ m.
(b) In section III we found that $L \sim 1$ mm if there are two extra dimensions. 
Thus the ADD model resolves the hierarchy $\ell_{{\rm P}} \ll \ell_{{\rm ew}}$ by introducing extra dimensions
of length scale $L$; however it suffers from a hierarchy problem of its own, namely, $\ell_{{\rm ew}} \ll L$.

{\em Problem 3. Three dimensional hyperbolic space.} Points in a flat four dimensional space can be labelled
by the four Cartesian co-ordinates $(x, y, z, w)$ that have the range $- \infty < x, y, z, w < \infty$. The space has
an indefinite metric. The distance between neighboring points is given by $ds^2 = dx^2 + dy^2 + dz^2 - dw^2$. 
The three dimensional hyperbolic space is defined as the set of points that 
satisfy the constraints $w^2 - z^2 - x^2 - y^2 = R^2$ and $w > 0$. Points on three dimensional hyperbolic space
can be labelled by the co-ordinates $(\psi, \theta, \varphi)$ that are related to the Cartesian co-ordinates via
\begin{eqnarray}
w & = & R \cosh \psi; \nonumber \\
z & = & R \sinh \psi \cos \theta; \nonumber \\
x & = & R \sinh \psi \sin \theta \cos \varphi; \nonumber \\
y & = & R \sinh \psi \sin \theta \sin \varphi.
\label{eq:3hyperboloid}
\end{eqnarray}
Here the co-ordinates have the ranges $0 \leq \psi < \infty$, $0 \leq \theta \leq \pi$ and $0 \leq \varphi < 2 \pi$.
(a) Justify the parametrization (\ref{eq:3hyperboloid}) and explain the ranges on the coordinates $(\psi, \theta, \varphi)$. 
(b) Determine the distance between two nearby points with co-ordinates $(\psi, \theta, \varphi)$ and $(\psi + d \psi, 
\theta + d \theta, \varphi + d \varphi)$. 

{\em Answer 3.} (a) The constraints that $w^2 - x^2 - y^2 - z^2  = R^2$ and $w > 0$ 
together imply that $w > R$. Hence we may write $w = R \cosh \psi$ with $0 \leq \psi < \infty$. 
Adopting this form the constraint then becomes $x^2 + y^2 + z^2 = R^2 \sinh^2 \psi$. 
From here we simply put down spherical polar co-ordinates on a sphere of radius $R \sinh \psi$. 
\newline
(b) $ds^2 = R^2 d \psi^2 + R^2 \sinh^2 \psi d \theta^2 + R^2 \sinh^2 \psi \sin^2 \theta d \varphi^2$. 

{\em Problem 4. DGP Analysis.} Here we fill in the steps that lead from eq (44) to eqs (48), (49) and (50). 
(a) Let $\tilde{g} (p\hspace{0mm})$ be the Fourier transform of $g(w)$.
Show that the Fourier transform of $g(w) \delta(w)$ is a constant given by
\begin{equation}
\int_{-\infty}^\infty \frac{dp}{2 \pi} \tilde{g}(p).
\label{eq:dgpfourier}
\end{equation}
Thus it follows that the Fourier transform of $- \delta(w) \nabla^2 \phi( x, y, z, w)$ is
$k^2 \tilde{f} ({\mathbf k})$ where 
\begin{equation}
\tilde{f} ({\mathbf k}) = \int_{-\infty}^\infty \frac{dp}{2 \pi} \; \tilde{\phi}({\mathbf k}, p)
\label{eq:dgpfourierii}
\end{equation}
and $\nabla^2 = \partial^2/\partial x^2 + \partial^2/\partial y^2 + \partial^2/\partial z^2$.
(b) Use the result of part (a) 
to rewrite eq (44) 
in Fourier space. You should obtain
\begin{equation}
\tilde{\phi} ({\mathbf k}, p) + \frac{\ell k^2}{k^2 + p^2} \tilde{f} ({\mathbf k}) = \frac{\lambda}{k^2 + p^2}.
\label{eq:dgpwavefourier}
\end{equation}
(c\hspace{0mm}) Use eqs (\ref{eq:dgpfourierii}) and  (\ref{eq:dgpwavefourier}) to determine $\tilde{f}$. You should obtain
\begin{equation}
\tilde{f} ({\mathbf k}) = \frac{\lambda}{2 k} \frac{1}{\left( 1 + \frac{1}{2} \ell k \right) }.
\label{eq:tildef}
\end{equation}
(d) Show that it is sufficient to know $\tilde{f}$ to determine $\phi({\mathbf r}, 0)$. One does not need 
$\tilde{\phi}$. You should find
\begin{equation}
\phi ({\mathbf r}, 0) = \int \frac{d {\mathbf k}}{(2 \pi)^3} \tilde{f}({\mathbf k}) \exp ( i {\mathbf k} \cdot {\mathbf r} ).
\label{eq:phidetermined}
\end{equation}
(e) The angular integrals in eq (\ref{eq:phidetermined}) can be exactly evaluated. Perform the angular integration.
You should obtain
\begin{equation}
\phi ({\mathbf r}, 0) = \frac{\lambda}{4 \pi^2 r} \int_0^\infty d k \; \frac{ \sin (k r) }{\left( 1 + \frac{1}{2} \ell k \right)}.
\label{eq:radialintegral}
\end{equation}
This is the exact expression for the potential of a point source in the DGP model. It can be rewritten in 
terms of rather obscure special functions (the cosine-integral and sine-integral functions) but those expressions are not
especially edifying. 
(f) Verify that eq (\ref{eq:radialintegral}) matches eqs (49) and (50) 
in the appropriate
limits. Useful asymptotic formulae: 
\begin{equation}
{\cal I}(\alpha) = \int_0^{\infty} dx \; \frac{\sin x}{1 + \alpha x}
\label{eq:asymptotics}
\end{equation}
has the asymptotic behavior ${\cal I} (\alpha) \approx 1 - 2! \alpha^2 + 4! \alpha^4 - 6! \alpha^6 + \ldots $
for small $\alpha$ and ${\cal I} (\alpha) \approx \pi/2 \alpha + [ \gamma_E - 1 + \ln (1/\alpha) ] (1/\alpha^2) + \ldots$
for large $\alpha$. Here $\gamma_E = 0.577216\ldots$ is Euler's constant. 

{\em Problem 5. RS1 Kaluza-Klein mode analysis.} 
(a) Verify that the zero-mode solution $\phi = 
g(\zeta, {\mathbf r}; {\mathbf k}) \exp ( i \omega t )$ satisfies the wave equation (21) as
well as the boundary condition $\partial \phi/\partial \zeta = 0$ at both $\zeta_l = \gamma^{-1}$ 
(left brane) and $\zeta_r = \gamma^{-1} \exp( \gamma \ell )$ (right brane). Here $g$ is given by eq (29).
(b) The Kaluza-Klein mode $\phi = f(\zeta, {\mathbf r}; {\mathbf k}) \exp( i \omega t)$, with $f$ given by 
eq (30) 
already satisfies the wave equation (21)
as well as the boundary condition on the left brane. Impose the boundary condition
$\partial \phi/\partial \zeta = 0$ on the right brane, $\zeta = \gamma^{-1} \exp ( \gamma \ell )$ to obtain the
quantization condition that must be satisfied by the allowed values of $\mu$. You should find
\begin{equation}
Y_1 \left( \frac{\mu}{\gamma} \right) J_1 \left( \frac{\mu}{\gamma} e^{\gamma \ell} \right) =
J_1 \left( \frac{\mu}{\gamma} \right) Y_1 \left( \frac{\mu}{\gamma} e^{\gamma \ell} \right).
\label{eq:kkquantization}
\end{equation}
(c\hspace{0mm}) Simplify eq (\ref{eq:kkquantization}) assuming that $(\mu/\gamma)e^{\gamma \ell} \gg 1$
and $\mu/\gamma \ll 1$. Use the resulting expression to determine approximate quantized values of $\mu$. 
You should obtain 
\begin{equation}
\cot \left( \frac{\mu}{\gamma} e^{\gamma \ell} - \frac{3}{4} \pi \right) \approx 0.
\label{eq:approximatequantization}
\end{equation}
for the approximate quantization condition and $\mu_n = \gamma \exp( - \gamma \ell ) ( n \pi + \frac{\pi}{4} )$
with $n = 0, 1, 2, \ldots$ for the quantized $\mu$ values. Rigorously the approximations made in this part are 
valid only for large $n$
but in fact these $\mu$ values are surprisingly accurate for all values of $n$. 

{\em Problem 6. Klein-Gordon field on RS1 right brane.} (a) Let us work with the original co-ordinates
$(t, x, y, z, w)$ on Randall-Sundrum space-time. In these co-ordinates the space-time interval is given 
by eq (19). 
As a prelude write the Klein-Gordon equation $\Box^2 \phi + \mu^2 \phi = 0$ for 
a scalar field $\phi$ that lives in the bulk. (b) Now let us consider scalar field that is confined to the right brane.
This scalar field is meant to represent a standard model particle. 
The space-time interval between neighboring points on the right brane is given by
\begin{equation}
ds^2 = e^{-2 \gamma \ell} (dt^2 - dx^2 - dy^2 - dz^2).
\label{eq:rightbranemetric}
\end{equation}
Use eq (\ref{eq:rightbranemetric}) to write the Klein-Gordon equation
$\Box_R^2 \xi + m^2 \xi = 0$ for a field $\xi$ confined to the right brane.
You should obtain
\begin{equation}
e^{2 \gamma \ell} \left( 
\frac{\partial^2}{\partial t^2} - \frac{\partial^2}{\partial x^2} - \frac{\partial^2}{\partial y^2}  - \frac{\partial^2}{\partial z^2} 
\right) \xi + m^2 \xi = 0.
\label{eq:rightbranekg}
\end{equation}
If we introduce the rescaled field $\overline{\xi} = \xi \exp( 2 \gamma \ell )$ we can bring the derivative terms in 
eq (\ref{eq:rightbranekg}) to the canonical form of a Klein-Gordon equation with mass parameter $m e^{-\gamma \ell}$. 
Choosing $m \hbar/c$ to be of order a Planck mass and choosing $\gamma \ell \approx $35-40 we can arrange
for standard model particles to have a mass of order 1 TeV, the electroweak scale, even though the
fundamental scale in the RS1 model is the Planck scale.

\end{document}